\documentclass[12pt]{article}

\def\bi{\begin{itemize}}
\def\ei{\end{itemize}}
\def\bea{\begin{eqnarray}} 
\def\eea{\end{eqnarray}}
\def\be{\begin{equation}}
\def\ee{\end{equation}}
\def\line{\hbox to \hsize}    
\def\frac #1#2{{#1\over #2}}

\def\tr{{\rm  tr\,}}

\def\bpsi{{\overline \psi}}

\def\n{{\bf n}}

\def\ket #1{{\vert #1\rangle}}
\def\bra #1{{\langle #1\vert}}
\def\dbrak#1#2{{\langle#1\vert #2\rangle}}
\def\brak #1#2{{\langle#1, #2\rangle}}
\def\eval #1#2#3{{\langle#1\vert#2\vert#3\rangle}} 
\def\vev #1{{\langle #1\rangle}}
\def\1{\mbox{\bf I}}

\def\bm#1{\mbox{\boldmath$#1$}} 

\newenvironment{Quote}
{\begin{list}{}{%
\setlength{\leftmargin}{10 pt}
\setlength{\rightmargin}{\leftmargin}}
\item[]}
{\end{list}}

     {\refstepcounter{exercisenumber} \begin{Quote}\small{\sl
     Exercise~\thechapter.\theexercisenumber\/}:}
{\end{Quote}}

{\begin{Quote}\small{\sl Example\/}:}
{\end{Quote}}

\def\levelonelist{
        \begin{list}{\mybulA}%
                        {
        \setlength{\topsep}{0pt}
        \setlength{\parsep}{0pt}
        \setlength{\partopsep}{0pt}
        \setlength{\itemsep}{0pt}
                        }
                }
\def\leveltwolist{
        \begin{list}{\mybulB}%
                        {
        \setlength{\topsep}{0pt}
        \setlength{\parsep}{0pt}
        \setlength{\partopsep}{0pt}
        \setlength{\itemsep}{0pt}
                        }
                }

\def\el{\end{list}}

\usepackage{amsfonts}   
\usepackage{amssymb}
\usepackage{centernot}
\usepackage[vcentermath]{youngtab}
\usepackage{appendix}
\usepackage{url}
\newcommand{\fsl}[1]{{\centernot{#1}}}

\def\aa{\accent23a} 

\begin{document}

\title{Gamma matrices,  Majorana fermions, and discrete symmetries  in Minkowski and Euclidean signature}

\author{ MICHAEL STONE\\
University of Illinois, Department of Physics\\ 1110 W. Green St.\\
Urbana, IL 61801 USA\\E-mail: m-stone5@uiuc.edu}


\maketitle

\begin{abstract}  I describe the interplay between Minkowski  and Euclidean signature 
gamma matrices, Majorana fermions, and discrete and continuous symmetries in all spacetime dimensions. \end{abstract}

\section{Introduction}
\label{SEC:introduction}

Textbook discussions of the discrete $C$, $T$  symmetries of the Dirac equation tend to feel   unsatisfactory because they make use of representation-specific properties of the  gamma matrices and other basis-dependent operations such as complex conjugation and  
transposition. Complex conjugation of the components of  a vector  is  basis-dependent  operation because the vector   can have real  components in one basis and imaginary  components in another, and when a matrix acts on a vector space $V$ its transpose acts on the dual space $V^*$. In the absence of some preferred  metric there is no natural  identification of a vector space with its dual  and   equating a spinor to a transposed spinor as we do in charge conjugation  is necessarily {\it unnatural}.    Other  linear-algebra aspects  of the Dirac equation --- the dimensions of the irreducible gamma-matrix representations,  Lorentz transformations,    and the existence of the Weyl spinors in even dimensions --- can be derived  purely from the Clifford algebra obeyed by the gamma matrices without using these operations. 
A particular consequence of the basis-choice dependence is that    Majorana fermions  are usually defined by  ugly equations containing non-covariant-looking products of gamma 
matrices that vary from book  to book \cite{curious}.

It  is also widely asserted that there is  no such thing as a Majorana fermion in four Euclidean dimensions. This  last is a great pity because we would like to study Majorana fermions using  heat-kernel  regularized  path integrals or by  lattice-theory computations,   and these tools are only available in Euclidean signature. 

The problem with this assertion is that many  authors identify the existence of Majorana fermions  with  the existence of real  gamma matrices and so, when the metric has  $p$ plus signs and $q$ negative signs, they need to delve into  the   reality properties of representations of the Clifford algebras ${\rm Cl}(p,q)$  . These   properties  depend intricately on $p-q$ (mod 8) \cite{polchinski,ofarril} and, in particular because 
$$
{\rm Cl}(p,q)\ncong {\rm Cl}(q,p),
$$
they suggest that the Bjorken and Drell  ``West Coast" or ``mostly minus" metric $g_{\mu\nu}= {\rm diag}(1,-1,-1,-1)$ is necessarily  different from  the ``East Coast," mostly plus,    metric  $g_{\mu\nu}= {\rm diag}(-1,+1,+1,+1)$  that  is preferred by general relativists.  Surely nothing { physical\/} can depend on our choice of  metric convention?

The representation theory of Clifford algebras defined over the field ${\mathbb R}$ contains much beautiful mathematics, but for physics   applications it rather misses the point. We compute  in Euclidean signature not because we are interested in  how physics would look  were our world Euclidean, but rather because it is a  useful tool for studying our  Minkowski-signature universe. 
Osterwalder and Schrader \cite{osterwalder-schrader} showed that when the Euclidean-signature  $n$-point functions satisfy the condition of reflection positivity then one can obtain  the  Minkowski-signature  $n$-point functions as analytic continuations of the Euclidean ones by taking the external momenta $p$  into  the Minkowski region where particles are on-shell when $p^2=-m^2$.
The {\it physical\/} Majorana condition in Euclidean signature does not need  real gamma matrices, but rather   that the reconstructed   $n$-point functions be those of Minkowski-signature Majorana particles \cite{wetterich}. The physical Majorana   condition  then depends only  the symmetry properties of certain matrices ${\mathcal C}_{\alpha\beta}$ and ${\mathcal T}_{\alpha\beta}$  which  are indifferent to the $\pm$ signs  in the metric \cite{murayama}.

A particular advantage of exploring the  properties of the ${\mathcal C}$ and ${\mathcal T}$ matrices is that they  make manifest an eightfold period in dimension that also appears in the classification of random matrices  \cite{dyson}, Cartan symmetric spaces, and in the classification of topological insulators and superconductors \cite{zirnbauer2,ryu-review}.  This  ``eightfold way"  is an aspect  of  of Bott periodicity \cite{bott,milnor} whose connection with the eightfold periodicity of gamma matrix properties    was pointed out by Ho{\v{r}}ava \cite{horava}  and Kitaev \cite{kitaev}.

Our aim is to use the ${\mathcal C}$ and ${\mathcal T}$ matrices to  reconcile the  problems stated above in  as straightforward a manner as possible, and to show how one can consistently switch between the field-operator language of Minkowski space and the Grassmann-variable path integral language in Euclidean signature. Our overall philosophy is similar that of Wetterich \cite{wetterich} although we stress different aspects of the mapping.

To achieve our  goal we  review in section \ref{SEC:gamma}  the construction of gamma matrices in $d$ spacetime dimensions and arbitrary signature  and in doing so uncover the properties of the ${\mathcal C}$ and ${\mathcal T}$  matrices that relate the gamma matrices  to their transpose. In section \ref{SEC:majorana} we will define the operation of charge conjugation in a basis and signature independent manner, and define three distinctly different types of Majorana fermions. We then how  the dimensions in which the three types  occur  --- $d=$ 2, 3, 4  (mod 8) for Majorana, $d=$8, 9, 10 (mod 8) for the necessarily massless pseudo-Majorana and $d=$5, 6, 7 (mod 8) for symplectic Majorana --- are the same in  both Minkowski and Euclidean signature.  We show how the period 8 pattern governing  both existence of the Majorana classes and their possible flavour and gauge symmetries is related to the eightfold version of Bott periodicity.  Section \ref{SEC:discrete} will then discuss how parity and time-reversal symmetries operate  in different dimensions.  Three appendices discuss how integrals over anticommuting variables differ between  Dirac and Majorana fermions (determinants {\it versus\/} Pfaffians) and also use the example of  condensed matter systems to show that  both   charge-conjugation  and time-reversal  act  antilinearly on the \underline{one}-particle space, but in the induced action on the  \underline{many}-particle Fock space charge conjugation acts linearly while     time-reversal  remains  antilinear.

\section{Gamma matrices} 
\label{SEC:gamma}

We begin with  the construction and properties of the Dirac gamma matrices in various dimensions and space-time signatures. This is well-known material and we will use the notation and strategy from   Hitoshi  Murayama's online lecture notes \cite{murayama}.

\subsection{\  ${\mathcal C}$ and ${\mathcal T}$ matrices: eightfold  periodicity}

In Euclidean   $d=2k$ dimensions we  can construct a matrix representation of the   generators $\gamma^i$ of the Clifford algebra 
\be
\gamma^i\gamma^j +\gamma^j\gamma^i =2\delta^{ij}
\label{EQ:clifford}
\ee 
by using  fermion annihilation and creation operators  $\hat a_n, \hat a^\dagger_n$, $n=1,\ldots, k$, that obey 
\be
\{\hat a_n,\hat a_m\}=0=\{\hat a^\dagger_n,\hat a^\dagger_m\}, \quad \{\hat a_n,\hat a^\dagger_m\}= \delta_{nm},
\ee
and by setting 
\bea
\gamma^{2n-1}&=&\hat a_n^{\dagger}+ \hat a_n, \nonumber\\
\gamma^{2n}&=& i(\hat a_n^\dagger-\hat a_n).
\eea
When  they   act  on the Fock space built on a   vacuum vector  $\ket{0}$ such that $\hat a_n\ket{0}=0$, $n=1,\ldots, k$, the $\gamma^i$  are represented by    a set of  $2^k$-by-$2^k$ Hermitian matrices  
that   are  symmetric for odd $i$  and antisymmetric for even $i$.

For odd dimension $d=2k+1$ we append an extra gamma matrix   
$$
\gamma^{2k+1} =  (-i)^k\gamma^1\ldots \gamma^{2k},
$$
which is equal to 
$ (-1)^{\sum \hat a^\dagger_n \hat a_n}$ and, in   this basis, is diagonal and so symmetric.

This     construction   
displays  a clear even-odd periodicity in the dimension $d$ because of the special treatment of $\gamma^{2k+1}$, but  it also contains a    rather less obvious period-eight property. 
 To reveal  this  hidden structure we  define  \cite{murayama}    
\be
C_1 = \prod_{i{\rm\, odd}} \gamma^i, \quad C_2= \prod_{i{\rm \,even}} \gamma^i,
\ee
and use them to construct  matrices ${\mathcal C}$ and ${\mathcal T}$ such that
\bea
{\mathcal C}\gamma^i {\mathcal C}^{-1} &=& - (\gamma^i)^T,\nonumber\\
{\mathcal T}\gamma^i {\mathcal T}^{-1} &=& + (\gamma^i)^T.
\eea

We find that   

\vskip 6pt 
\noindent {\bf k=0, mod 4}: 
${\mathcal C}=C_1$  symmetric, ${\mathcal T}=C_2$  symmetric. \\Both commute with $\gamma^{2k+1}$. 

\vskip 6pt 
\noindent {\bf k=1, mod 4}:
${\mathcal C}=C_2$ antisymmetric,  ${\mathcal T}=C_1$ symmetric.\\Both \underline{anti}commute  with $\gamma^{2k+1}$.

\vskip 6pt 
\noindent {\bf k=2, mod 4}:
${\mathcal C}=C_1$  antisymmetric,  ${\mathcal T}=C_2$ antisymmetric  \\Both commute  with $\gamma^{2k+1}$.

\vskip 6pt 
\noindent {\bf k=3, mod 4}:
${\mathcal C}=C_2$  symmetric,  ${\mathcal T}=C_1$  antisymmetric. \\Both \underline{anti}commute  with $\gamma^{2k+1}$.

Under a change of basis $\gamma^\mu \to A\gamma^\mu A^{-1}$  the  matrices    ${\mathcal C}$ and ${\mathcal T}$ will  no longer given by the explicit product expressions  $C_1$ and $C_2$, but  instead  transform as  
\be
{\mathcal C} \to A^T {\mathcal C} A,\quad {\mathcal T} \to A^T {\mathcal T} A.
\ee
The symmetry or antisymmetry  of  ${\mathcal C}$, ${\mathcal T}$ is  unchanged,  and  is thus  a basis-independent property

  Another way to think of this symmetry  is   by making use of the transpose of   their defining transformations to see   that ${\mathcal C}^{-1}{\mathcal C}^T$ and ${\mathcal T}^{-1}{\mathcal T}^T$ commute with all $\gamma^\mu$. As our  Fock-space gamma  representation is clearly   irreducible,  Schur's lemma tells  us that both  ${\mathcal C}, {\mathcal T}$ are proportional to their transpose 
so
\be {\mathcal C}^T =\lambda {\mathcal C}
\ee
with $\lambda$  basis independent. 
Then, transposing again, 
\be
{\mathcal C}= \lambda {\mathcal C}^T \Rightarrow  {\mathcal C}= \lambda^2 {\mathcal C}
\ee
showing that   $\lambda=\pm 1$. Similarly ${\mathcal T}^T=\pm {\mathcal T}$ with a basis independent sign.

If we restrict to transformations in which $A$ is unitary, the Euclidean $\gamma^\mu$ remain  Hermitian and  a similar argument shows that ${\mathcal C}^\dagger {\mathcal C}$ is proportional to the identity.  As  ${\mathcal C}^\dagger {\mathcal C}$ is a positive operator the factor of proportionality  is real and positive.  Consequently  ${\mathcal C}$ (and  ${\mathcal T}$) can be scaled  by real numbers so as to be   unitary. We will assume that we have done this.

If  we regard a   gamma  matrix with  elements ${\gamma^\alpha}_\beta$ as representing  a linear map from $V\to V$, where $V$ is the spinor representation space, then its transpose $\gamma^T$ with matrix elements ${(\gamma^T)_\beta}^\alpha$ 
  represents a linear map from $V^*\to V^*$ where $V^*$ is the dual space of $V$. We can think of the matrices 
${\mathcal C}_{\alpha\beta}$ and ${\mathcal T}_{\alpha\beta}$, and their inverses  $({\mathcal C}^{-1})^{\alpha\beta}$ and $({\mathcal T}^{-1})^{\alpha\beta}$, as ``metrics'' on spinor-space that  allow us to raise and lower the spinor indices on the Fermi fields  $\psi^\alpha$ and $\bar\psi_\beta$ and so identify $V$ with $V^*$. The  existence of ${\mathcal C}_{\alpha\beta}$ and ${\mathcal T}_{\alpha\beta}$ thus    resolves   one of the unnaturalness issues  raised in the introduction.  It is then clear that a matrix product such as ${\mathcal C}\gamma^i$ is well formed and covariant  because  an upstairs index on the gamma matrix is  contracted  with a downstairs index on ${\mathcal C}$. A matrix product such as  $\gamma^i {\mathcal C}$ is {\it not\/}  well-formed as we are contracting a pair of downstairs indices.  These observations can serve as a useful check on manipulations.

In a similar vein one sometimes sees   assertions such as   ``${\mathcal T}^2={\mathbb I}$''   (implicitly in Bjorken and Drell \cite{bjorken} eq.\ (15.134) for example)  but the manner in which   ${\mathcal T}$ and ${\mathcal C}$ transform under $\gamma^i \to A\gamma^i A^{-1}$ shows that there is no basis-independent notion of a product of ${\mathcal T}$ or  ${\mathcal C}$ matrices with themselves.  Such   
formul\ae\  are  not operator identities therefore, and can only hold in specific bases. There is no such problem with ${\mathcal C}^{-1}{\mathcal C}^T=\pm {\mathbb I}$, {\it etc\/}.

If the matrices $\gamma^i$ constitute a representation of the Clifford algebra, so do $\pm (\gamma^i)^T$.  In $d=2k$  dimensions the Dirac representation is unique, so these three representations must be equivalent. Consequently, even if we did not have the explicit construction given above,  the existence of  $\mathcal T$, $\mathcal C$  is guaranteed.
In {\it odd\/} dimensions, however, there are {\it two\/} inequivalent representations of the Clifford  algebra because we can replace   $ \gamma^{2k+1}=  (-i)^k\gamma^1\cdots \gamma^{2k}$ by minus this expression and still satisfy the Clifford algebra. The  representation with the minus sign is inequivalent to that with the plus sign because $\gamma^1\gamma^2\cdots \gamma^{2k+1} = \pm {\mathbb I}$ and the sign cannot be changed by any transformation $\gamma^\mu \to U^{-1} \gamma^\mu U$. The existence of  $\mathcal T$, $\mathcal C$ matrices that transpose {all} $2k+1$  $\gamma^i$ is no therefore longer assured. We have to ask   whether  conjugation by the  even-dimensional   $\mathcal T$, $\mathcal C$  continues to correctly transpose  the extra gamma matrix.  Table  \ref{TAB:conjugation} summarizes the outcome of this examination. 
\begin{table}
\begin{center}
\begin{tabular}{|c|ccccccccccc|}
\hline\hline
d&0&1&2&3&4&5&6&7&8&9&10\\
\hline
${\mathcal T}$  &S&S&S& &A&A&A& &S&S&S\\
${\mathcal C}$   &S& &A&A&A& &S&S&S& &A\\
${\gamma^{d+1}}$  &+& &$-$& &+ & &$-$ & &+&&$-$\\
\hline\hline
\end{tabular}
\end{center}
\caption{\sl Dimension dependence of $\mathcal T$ and  $\mathcal C$ . The table  shows whether  the $\mathcal T$ and  $\mathcal C$  matrices exist,  whether they are symmetric (S), or antisymmetric (A), and, for $d$ even,  the sign appearing  in      
$
{\mathcal C}\gamma^{d+1}{\mathcal C}^{-1}= {\mathcal T}\gamma^{d+1}{\mathcal T}^{-1}= \pm (\gamma^{d+1})^T$.
 The table repeats mod 8. This is an aspect of Bott periodicity \cite{kitaev}. }
\label{TAB:conjugation}
\end{table}

When ${\mathcal T}$ is  symmetric we can find a unitary matrix $U$ with which to transform  
$
{\mathcal T}\to {\mathcal T}'=U^T {\mathcal T}U$  to  
a basis in which ${\mathcal T}'= {\mathbb I}$ (see appendix \ref{SEC:canonical} for a proof of this).  In this basis  all the Euclidean gamma matrices are  symmetric, still Hermitian, and therefore all real. When ${\mathcal C}$ is symmetric we can find a basis in which ${\mathcal C}=\mathbb I$ and all the  Euclidean gamma matrices are  {\it antisymmetric\/}, still Hermitian, and therefore  purely {imaginary}.

In other signatures the $\delta^{\mu\nu}$ in the defining equation (\ref{EQ:clifford}) will be replaced by ${\rm diag}(\pm 1,\pm 1 ,\ldots,\pm 1)$ and for each minus  sign the corresponding $\gamma^\mu$ must be multipled by $\pm i$ in order to satisfy the new Clifford algebra. The reality properties of the gamma matrices are obviously changed by this. One    advantage of focussing on the ${\mathcal C}$ and ${\mathcal T}$ matrices   is that their existence and symmetry properties are indifferent to the metric signature. 

\subsection{ Numbering convention for Minkowski-signature matrices} 
We labelled  our Euclidean gamma matrices as $\gamma^1,\gamma^2, \ldots, \gamma^{2k}$ with  $\gamma^{2k+1}$ being a product of the $2k$ lower-numbered matrices. In contemporary physics usage four-dimensional Minkowski-signature  gamma matrices  are universally numbered as $\gamma^0, \gamma^1, \gamma^2, \gamma^3$ with $\gamma^0$ associated with $x^0=t$. For historical reasons their   product still  called $\gamma^5$ --- although there is  no $\gamma^4$.  In Minkowski signature $\gamma^0$ and $\gamma^5$ have special roles and     renaming either   $\gamma^0\to \gamma^1$ or $\gamma^5\to \gamma^4$   to close  the ``$\gamma^4$ gap'' 
  is likely to generate  more fog  than light. It seems  simplest to keep the chirality operator  as $\Gamma^5\equiv \gamma^{2k+1}$,  and  when  ``$\gamma^0$" appears in  the familiar definition   $\bar\psi = \psi^\dagger\gamma^0$  it should be born mind that the  Minkowski signature  ``$\gamma^0$" corresponds to the Euclidean   $\gamma^4$.

\section{Charge conjugation and Majorana fermions}
\label{SEC:majorana}

\subsection{Charge conjugation}

In a  Euclidean-signature path integral the Fermi fields $\bar \psi$ and $\psi$ are unrelated   row and column vectors of  Grassmann variables (see Appendix \ref{SEC:berezin}).   Nonetheless it is useful to  define the   Euclidean-signature charge-conjugate fields $\bar \psi^c$ and $\psi^c$  so as to  be consistent with the Minkowski-signature operator language  in which $\bar \psi$ and $\psi$ {\it are\/} related by $\bar\psi =\psi^\dagger \gamma^0$. We arrange for this    by  defining 
\bea
\psi^c&=& {\mathcal C}^{-1}{\bar \psi}^T,\nonumber\\
\bar \psi^c &=& -\psi^T {\mathcal C}.
\eea

To obtain  the motivating Minkowski  version  we recall  that in any signature we can use exactly the same ${\mathcal T}$ and ${\mathcal C}$ matrices  as in Euclidean signature --- the insertion of factors of $i$ in some of the  $\gamma^\mu$ does not  affect  the formula for their transposition  and  no $i$'s need be inserted in ${\mathcal T}$ and ${\mathcal C}$.

Consider first   the  mostly-minus ``West-Coast'' Minkowski metric  $(+,-,-,\ldots)$ in which    $\gamma^0$  is Hermitian and obeys $(\gamma^0)^2=1$. Then with $\bar\psi^T = (\psi^\dagger \gamma^0)^T $, and writing $\psi^*$ for the quantum Hilbert-space adjoint of $\psi$ without  the column $\to $ row operation implicit in $\psi^\dagger$,  we have 
\be
\psi^c= {\mathcal C}^{-1}{\bar \psi}^T=   {\mathcal C}^{-1} (\gamma^0)^T \psi^*\,  \Rightarrow (\psi^c)^\dagger = \psi ^T(\gamma^0)^T {\mathcal C}
\ee
because    ${\mathcal C}$ remains  unitary  in Minkowski space. We define  
\bea
\bar \psi^c 
&\equiv& \overline{(\psi^c)}\nonumber\\
&=& (\psi^c)^\dagger  \gamma^0\nonumber\\
 &=& \psi^T (\gamma^0)^T {\mathcal C} \gamma^0\nonumber\\
&=&- \psi^T \,{\mathcal C} \gamma^0 {\mathcal C}^{-1}{\mathcal C} \gamma^0\nonumber\\
&=& -\psi^T \,{\mathcal C}.
\eea

In the mostly-plus ``East-Coast'' Minkowski metric $(-,+,+,\ldots)$, in which $\gamma^0$ is skew Hermitian and obeys $(\gamma^0)^2=-1$,  we have $(\psi^c)^\dagger= -\psi ^T(\gamma^0)^T {\mathcal C}$ and 
\bea
 \bar \psi^c &\equiv& \overline{(\psi^c)}\nonumber\\
  &=& (\psi^c)^\dagger  \gamma^0\nonumber\\
 &=&- \psi^T (\gamma^0)^T {\mathcal C} \gamma^0\nonumber\\
&=& \psi^T \,{\mathcal C} \gamma^0 {\mathcal C}^{-1}{\mathcal C} \gamma^0\nonumber\\
&=& \psi^T \,{\mathcal C}(\gamma^0)^2\nonumber\\
&=& -\psi^T \,{\mathcal C}.
\eea
In both signatures, therefore, $\bar \psi^c = -\psi^T \,{\mathcal C}$.

From these results, and with anticommuting Grassmann  $\psi$'s,  we find that
\be
\bar \psi^c \gamma^\mu\psi^c =  [-\psi^T {\mathcal C}]\gamma^\mu [{\mathcal C}^{-1} \bar \psi^T]= \psi^T (\gamma^\mu)^T\bar \psi^T = - \bar \psi \gamma^\mu\psi,
\ee
so the number current changes sign.
The spin-current  
density  transforms as 
\be
\bar \psi^c \gamma^0 [\gamma^i,\gamma^j]\psi^c= -\bar\psi \gamma^0 [\gamma^j, \gamma^i]\psi = \bar \psi \gamma^0 [\gamma^i,\gamma^j]\psi, \quad (i,j\ne 0),
\ee
and is left unchanged. 
Similarly
\be
\bar \psi^c \psi^c =- \psi^T \bar \psi^T = \bar\psi\psi.
\ee
In Euclidean signature, and using the  anticommuting property of the Grassmann fields,  the action for Dirac fermions minimally-coupled to a  skew-Hermitian vector gauge field $A_\mu$ has the property
\be
S= \int d^n x\, \bpsi [\gamma^i (\partial_i+A_i)+m]\psi =  \int d^n x\, {\bar\psi}^c [\gamma^i (\partial_i-A^T_i)+m]\psi^c.
\ee
The  $-A^T$ are the Lie algebra representation-valued fields in the the conjugate representation to that of $A$, and so $\psi^c$ has  the opposite gauge-field ``charge'' to $\psi$.

\subsection{ Minkowski-signature Majorana Fermions}

 We have defined 
\be
\psi^c= {\mathcal C}^{-1} {\bar \psi}^T={\mathcal C}^{-1} (\gamma^0)^T \psi^*
\ee
so, with ${\mathcal C}^T=\lambda {\mathcal C}$ we find (in both mostly-plus and mostly-minus metrics)
\bea
(\psi^c)^c&=&  {\mathcal C}^{-1} (\gamma^0)^T ({\mathcal C}^{-1} (\gamma^0)^T\psi^*)^*\nonumber\\
&=& {\mathcal C}^{-1}(\gamma^0)^T {\mathcal C}^T(\gamma^0)^\dagger \psi\nonumber\\
&=& \lambda \,{\mathcal C}^{-1}(\gamma^0)^T {\mathcal C}(\gamma^0)^\dagger \psi\nonumber\\
&=& -\lambda \gamma^0 (\gamma^0)^\dagger \psi\nonumber\\
&=& -\lambda \psi.
\eea
We can therefore consistently impose the  Minkowski Majorana condition that $\psi^c=\psi$ only if $\lambda=-1$ so  ${\mathcal C}$ is {\it antisymmetric\/}: {\it i.e.\/}\ in 2, 3, 4 (mod 8) dimensions.  

The equal-time anti-commutator of an operator-valued  Majorana field can be taken to be  
\be
\{\psi^\alpha(x),\psi^\beta(x')\}_{t=t'}= [\gamma^0 {\mathcal C}^{-1}]^{\alpha\beta} \delta^{d-1}(x-x')
\ee
where  $\gamma^0{\mathcal C}^{-1}= -{\mathcal C}^{-1} (\gamma^0)^T$ is symmetric when ${\mathcal C}$ is antisymmetric.

We can  regard the map 
${\textsf C}:\psi\mapsto \psi^c= {\mathcal C}^{-1} [\gamma^0]^T \psi^*$ as an antilinear\footnote{Charge conjugation is antilinear only when acting on the field components. It is a \underline{linear} map when acting on  the states in the many-body Hilbert space. See Appendix \ref{SEC:condensed}.}    map ${\textsf C}: V \to V $ where $V$ is the gamma-matrix representation space. If ${\textsf C}^2={\rm id}$, this is {\it real structure\/}  on the complex   $V$ space. Vectors that are left fixed by $\textsf C $ are regarded ``real''  because  there is a basis in which their components are real --- even even though these components will  be complex in other bases.

The antilinear map $\textsf C$ commutes with the gamma matrices only in the { mostly plus\/} East Coast metric. With this metric choice, and in the basis in which the Majorana spinor components  are real,   the gamma matrices become   purely real and so preserve the reality condition.  In the the West-Coast-metric  Majorana representation the gamma's are purely imaginary and we have to remove a factor of $i$ to get matrices that commute with the antilinear $\textsf C$. This does not  matter though,  because it is  the {\it Dirac equation\/}  that must preserve the reality of of the spinor solutions, and in the West Coast Minkowski metric the Dirac equation    is
\be  
(-i \gamma^i\partial_i+m)\psi=0\quad \hbox{(West Coast)}.
\ee
This version of the equation   puts  the necessary  factor of $i$ with the $\gamma$'s, while  on the East Coast the  Dirac equation reads
\be
( \gamma^i\partial_i +m)\psi=0,\,\,\,\quad \hbox{(East Coast)},
\ee
where  there is no factor of $i$.

To  verify that $\textsf C$ commutes with the $\gamma^i$ in the East Coast metric we begin  by observing  that $(\gamma^i)^\dagger = \gamma^0 \gamma^i\gamma^0$ in both conventions. 
Then
\bea
{\mathcal C}^{-1} (\gamma^0)^T(\gamma^{i} \psi)^*&=&  {\mathcal C}^{-1} (\gamma^0)^T(\gamma^{i*} \psi^*)\nonumber\\
&=&  {\mathcal C}^{-1}( \gamma^0)^T(\gamma^0\gamma^i \gamma^0)^T \psi^*\nonumber\\
&=&{\mathcal C}^{-1} (\gamma^0)^T(\gamma^0)^T (\gamma^i)^T (\gamma^0)^T \psi^*\nonumber\\
&=&{\mathcal C}^{-1} (\gamma^0)^T{\mathcal C}{\mathcal C}^{-1} (\gamma^0)^T {\mathcal C}{\mathcal C}^{-1}(\gamma^i)^T{\mathcal C}{\mathcal C}^{-1} (\gamma^0)^T \psi^*\nonumber\\
&=& (-\gamma^0)(-\gamma^0)(-\gamma^i) {\mathcal C}^{-1} (\gamma^0)^T \psi^*\nonumber\\
&=& -(\gamma^0)^2\gamma^i  {\mathcal C}^{-1} (\gamma^0)^T\psi^*.
\eea
Thus $\textsf C \gamma^i = \gamma^i \textsf  C  $, or equivalently  
\be
{\mathcal C}^{-1} (\gamma^0)^T(\gamma^{i} \psi)^*{=} \gamma^i  \,{\mathcal C}^{-1} (\gamma^0)^T\psi^*
\ee
holds only if $(\gamma^0)^2=-1$.

\subsection{Minkowski-signature pseudo-Majorana fermions}

We can define an alternative  ``charge conjugation''  operation  
\bea
\psi^\tau&=& {\mathcal T}^{-1}{\bar \psi}^T,\nonumber\\
\bar \psi^\tau &=& \psi^T {\mathcal T}.
\eea
This operation reverses the current, again leaves the spin unchanged, but flips the sign of $\bar\psi \psi$.  Almost identical algebra to the conventional  charge  conjugation case shows that the condition $\psi^\tau=\psi$ is consistent only when ${\mathcal T}$ is {\it symmetric\/},  hence in $d$= 8, 9, 10  (mod 8). 
 Fermions  such that $\psi^\tau=\psi$ are  said by some authors  \cite{kugo,tanii,freund,nieuwenhuizen} to be 
 {\it pseudo-Majorana\/}\footnote{ 
 Jos{\'e}  Figueroa-O'Farril \cite{ofarril} also uses the  term  {pseudo-Majorana\/} spinors, (``a nebulous concept best kept undisturbed")  but by this I believe  he means the   purely imaginary gamma matrices of the West-Coast Majorana representation.
 }.  
  
  Repeating the algebra for the $\textsf  C$ conjugation, but with ${\mathcal C}$ replaced by $\mathcal T$, gives an extra minus sign. Consequently  in the  mostly-minus West-Coast metric the gamma matrices of  a  pseudo-Majorana representation can be chosen to be \underline{real}, while in a Majorana representation they are pure imaginary. It is the other way around in the mostly-plus East-Coast metric. 

 Because this ``conjugation''  flips $\bar\psi\psi$, these pseudo-Majorana fermions are necessarily massless. Indeed  the absence of the mass term is necessary for   the  real gamma matrices  in the   West Coast  pseudo-Majorana representation and the pure imaginary gamma matrices in the East Coast pseudo-Majorana representation   to avoid     conflict  with their appropriate Dirac equation.
 

\subsection{Euclidean-signature Majorana fermions}
 We now  explore to what extent  the  constraints on the Minkowski signature space-time dimensions in which Majorana and pseudo-Majorana fermions  exist are compatible  with  Euclidean-signature Grassmann-variable path integration.

Assume that any   gauge fields in the skew-Hermitian Euclidean-signature Dirac operator 
\be
{\fsl D}= \gamma^\mu (\partial_\mu+A_\mu)
\ee
are in real representations so that $A_\mu$ is a real matrix and $A_\mu=-A_\mu^T$.
Then, if we have an eigenfunction such that  
\be
{\fsl D}u_n=i\lambda_n u_n,
\ee
complex conjugation gives    
\be
{\fsl D}^* u^*=-i\lambda_n u^*_n. 
\ee
This  can be written as  
\be
{\mathcal C} {\fsl D} {\mathcal C}^{-1} u^*_n = i\lambda_n u^*_n,
\ee
or 
\be
{\fsl D} {\mathcal C}^{-1} u^*_n = i\lambda_n  {\mathcal C}^{-1} u^*_n.
\ee
Thus $u_n$ and ${\mathcal C}^{-1} u_n^*$ are both  eigenfunctions of ${\fsl D}$  with the same eigenvalue\footnote{We do not really need the complex conjugation. If   we  do not  demand that  the $A$ in $\gamma^\mu\to A\gamma^\mu A^{-1} $   be unitary our  Euclidean gamma-matrices are no longer  Hermitian but we  can still obtain the second eigenfunction for a given $\lambda$ as $[{\mathcal C}^{-1}]^{\alpha\beta} v_\beta $ where $v_\beta  $ is a {\it left eigenvector\/} of ${\fsl D}$; {\it i.e.\/}\ one that  obeys  $v\overleftarrow {\fsl D}=i\lambda v$ where $v\overleftarrow{\fsl D}\equiv (-\partial_\mu v) \gamma^\mu$.  See appendix  \ref{SEC:berezin}  eq \ref{EQ:left-right-eigenvector}  {\it et seq\/}. } .  They will be orthogonal, and therefore linearly independent, when ${\mathcal C}$ is antisymmetric --- something that happens in  $d=2,3,4$ (mod 8)  Euclidean dimensions. These are precisely the dimensions in which {\it Minkowski space\/} Majorana spinors can occur.  
This suggests that we can take the  
 Euclidean Majorana-Dirac action to be 
\be
S[\psi]= \frac 12 \int d^dx \,\psi^T {\mathcal C}( {\fsl D}+m)\psi.
\ee
The combination   $\psi^T{\mathcal  C}$ is called by Peter van Nieuwenhuizen \cite{nieuwenhuizen}  the ``Majorana adjoint." 

Expanding the  fields out as 
\bea
\psi (x) &=& \sum_n[ \xi_n u_{n}(x)+ \eta_n( {\mathcal C}^{-1}u_{n}^*(x))], \nonumber\\
\psi^T(x)   &=& \sum_n [\xi_n u^T_n(x) - \eta_n (u_n^\dagger(x) {\mathcal C}^{-1})],
\eea
where $\xi_n$ and $\eta_n$ are Grassmann variables
 we find that  
\bea
S&=&    \frac 12 \int d^dx \sum_{i,j} ( \xi_i u^T_i - \eta_i u_i^\dagger {\mathcal C}^{-1}) {\mathcal C}(i\lambda_j+m)( \xi_j u_j+ \eta_j {\mathcal C}^{-1}u_j^*)\nonumber\\
&=&  \frac 12 \int d^dx  \sum_{i, j}\left\{ \xi_i\eta_j (u^T_i u_j^*) + \xi_j \eta_i (u_i^\dagger u_j)\right\}(i\lambda_j+m) \nonumber\\
&=& \sum_i \xi_i\eta_i (i\lambda_i+m).
\eea 
As the Grassmann integration uses  only one copy of the doubly degenerate eigenvalue, we obtain  a square-root of the full Dirac determinant \hbox{${\rm Det}({\fsl D}+m)$.}   
We anticipate  that the resulting partition function is the Pfaffian (see Appendix \ref{SEC:pfaffian})
\be
Z= {\rm Pf}[ {\mathcal C}( {\fsl D}+m)]
\ee
of the skew-symmetric bilinear kernel     ${\mathcal C}( {\fsl D}+m)$. 

To confirm  that the kernel   is skew symmetric  we take  a transpose and   find 
\bea  
[{\mathcal C}( \gamma^\mu \partial_\mu +m)]^T &=&  (\partial^T_\mu (\gamma^\mu)^T+m){\mathcal C}^T\nonumber\\
&=&[(- \partial_\mu) (-{\mathcal C}\gamma^\mu {\mathcal C}^{-1}) +m] (-{\mathcal C})\nonumber\\
&=& -{\mathcal C}(\gamma^\mu \partial_\mu+m).\
\eea 
Note that  $\partial^T=-\partial$ because its   $x$-basis matrix element is $\delta'(x-x')$ is  skew. The skew symmetry  continues to  hold for ${\fsl D}$ in curved space and  with   gauge fields in real representations. 

One  further step is needed to confirm that
\be
{\rm Pf}[ {\mathcal C}( {\fsl D}+m)]=  \prod_n (i\lambda_n+m).
\ee
We know  from Appendix \ref{SEC:pf-proof} that under a  change of basis a  Pfaffian transforms as ${\rm Pf} [{B}^T{ QB}]=  {\rm Pf}[{ Q}] \,{\rm det}[{B}]$. The mode  expansion
\be
 \psi^\alpha (x) = \sum_n[ \xi_n u^{\alpha n}(x)+ \eta_n( [{\mathcal C}^{-1}]^{\alpha\beta}u_{\beta n}^*(x))],
 \ee
is a such linear change of variables  in the Grassmann integral. It  corresponds to a matrix ${ B}$ with indices $B_{\alpha n ; \beta x}$, where the range of $n$ is doubled to  include both the labels on $u_n$ and on  ${\mathcal C}^{-1}u_{n}^*$. To show  that diagonalizing ${\fsl D}$ has not altered  the value of the Pfaffian we need to show that this matrix is unimodular --- or at least does not depend on the background fields. However  
\be
\int d^dx \frac 12 \psi^T(x) {\mathcal C}\psi(x) = \sum_n \xi_n\eta_n= \frac 12 \sum_i (\xi_n\eta_n-\eta_n \xi_n)
\ee
holds for any set of  eigenmodes $u_n$.  In other words 
\be
{B}^T ({\mathcal C}\otimes {\mathbb I}) {B}= {\mathcal C} \otimes \tilde {\mathbb I}
\ee
where $[{\mathcal C}\otimes {\mathbb I}]_{\alpha n; \beta m}= {\mathcal C}_{\alpha\beta}\delta_{nm}$ and  $[{\mathcal C} \otimes \tilde {\mathbb I}]_{\alpha x;\beta x'}= { \mathcal C}_{\alpha\beta} \delta^d(x-x')$,
 The matrices  ${B}$ from different background fields  can differ only by left factors that preserve the symplectic form $ ({\mathcal C}\otimes {\mathbb I})$. Such symplectic matrices are automatically unimodular.
 
 We could redefine  the mode expansion  as in \cite{golterman} 
 \be \psi^\alpha (x) = \sum_n[ \xi_n u^{\alpha n}(x)+ \eta_ne^{i\theta_n} ( [{\mathcal C}^{-1}]^{\alpha\beta}u_{\beta n}^*(x))],
\ee
where the $e^{i\theta_n}$ are arbitrary phases
This modification   replaces $(i\lambda_n+m)\to e^{i\theta_n}(i\lambda_n+m)$ in the eigenvalue product and apparently alters the Pfaffian. The $B$ matrix is no longer  unimodular, though, and its phase  cancels the product of the  phase   factors in the modified eigenvalue product leaving  ${\rm Pf}[{\mathcal C}({\fsl D}+m)]$  unaffected. We therefore disagree  with the claim in \cite{golterman}  that the phase of the Pfaffian is ill-defined.

 \subsubsection{\bf 4-d   Dirac {\it vs.}\  Majorana}
 
 As a further illustration of the need to be careful when performing  linear transformations on the Pfaffian integrands  we consider the pulling apart a Dirac  field  into two Majorana fields.
   In  four dimensions we can do this by relabelling  its  Weyl fermion components  as\footnote{If we decompose a Dirac spinor into it's Weyl-fermion chiral parts $\psi=[\psi_R,\psi_L]^T$ then in the Minkowski operator language we have $\bar \psi=[\psi_L^\dagger,\psi_R^\dagger]$. It is a common convention to label  the entries in $\bar \psi$ so that $\psi_R^\dagger\to  \bar\psi_R$ and $\psi_L^\dagger\to \bar\psi_L$ making $\bar\psi=[\bar\psi_L,\bar\psi_R]$.}
\bea
\psi\quad=\quad \left[\matrix{\psi_R \cr \psi_L}\right]&=&\left[\matrix{ \chi_{2R}\cr
\chi_{1L}}\right]\nonumber\\
\bar\psi\quad =\,[\bar\psi_L,  \bar\psi_R] &=&[-\chi^T_{1R}, 
-\chi^T_{2L}]{\mathcal C}.
\eea
If all gauge fields are in real representations we can rewrite the kinetic part of the action density 
\be
{\mathcal L}=[\bar\psi_L,\bar\psi_R]\left[\matrix{0& {\fsl D}_L\cr {\fsl D}_R&0}\right] \left[\matrix{\psi_R\cr \psi_L}\right]
\ee
as  
\bea
{\mathcal L} &=&-\frac 12\left\{ [\chi^T_{1R} ,\chi^T_{2L}]{\mathcal C} \left[\matrix{0& {\fsl D}_L\cr {\fsl D}_R&0}\right] \left[\matrix{\chi_{2R} \cr \chi_{1L}}\right]
+  [\chi^T_{2R},\chi^T_{1L}]{\mathcal C}\left[\matrix{0& {\fsl D}_L\cr {\fsl D}_R&0}\right] \left[\matrix{\chi_{1R} \cr \chi_{2L}}\right]\right\}\nonumber\\
&=& -\frac 12\left\{ [\chi^T_{1R} ,\chi^T_{1L}]{\mathcal C}\left[\matrix{0& {\fsl D}_L\cr {\fsl D}_R&0}\right] \left[\matrix{\chi_{1R} \cr \chi_{1L}}\right]
+  [\chi^T_{2R} ,\chi^T_{2L}]{\mathcal C}\left[\matrix{0& {\fsl D}_L\cr {\fsl D}_R&0}\right] \left[\matrix{\chi_{2R} \cr \chi_{2L}}\right]\right\}\nonumber\\
&=& \frac 12 \left\{\bar \chi_1 {\fsl D}\chi_1 + \bar \chi_2 {\fsl D}\chi_2 \right\}.
\eea
Here $\bar\chi=[\bar \chi_L, \bar\chi_R]  = [-\chi^T_R , -\chi^T_L ]{\mathcal C}$ and both $\chi_1$ and $\chi_2$ are Majorana because $\bar\chi= -\chi^T{\mathcal C}$. The averaging in the first line comes from antisymmetry of ${\mathcal C}{\fsl D}$.
A Dirac mass term 
\be
m \bar \psi\psi= m[\bar\psi_L,\bar\psi_R]\left[\matrix{\psi_R\cr \psi_L}\right]
\ee
becomes 
\be
\frac{m}{2}\left\{  \bar \chi_1 \chi_2+  \bar \chi_2 \chi_1\right\}= \frac m2 \bar \chi \sigma_1 \chi
\ee
where the $\sigma_1$ acts on the ``flavour'' indices 1,2. 
We can do a flavour diagonalization by 
\bea
\chi&\to& e^{i\pi \sigma_1 \gamma^5/4 } \chi = \frac 1{\sqrt{2}} (1+i\sigma_1\gamma^5)\chi\nonumber\\
\bar\chi &\to& \bar\chi e^{i\pi \sigma_1 \gamma^{5}/4}= \bar\chi \frac 1{\sqrt{2}} (1+i\sigma_1\gamma^5)
\eea
that takes 
\be
 \bar \chi \sigma_1 \chi\to \bar\chi \sigma_1  e^{i\pi \sigma_1 \gamma^5/2} \chi =\bar\chi i (\sigma_1)^2 \gamma^5\chi
\ee
to get
\be
\frac m 2 \left\{\bar\chi_1 (i\gamma^5)\chi_1+ \bar\chi_2 (i\gamma^5)\chi_2\right\}.
\ee
The transformation is unimodular for both eigenvalues of $\gamma^5$, and so it seems as if the path integral outputs  the product of  two Pfaffians (and therefore a determinant) of fields with an $im \gamma^5$  chiral mass. As every eigenvalue occurs twice this means that product of the two identical Pfaffians, which should reproduce the Dirac determinant, has become 
\be
|m|^{n_++n_-} e^{i(\pi/2)(n_+-n_-)}\prod_n(m^2+\lambda_n^2).
\ee
Here $n_+$ and $n_-$ are the number of zero modes with plus or minus $\Gamma^5$ chiralities.
This determinant  has  apparently acquired a  factor of $(-1)$ for each pair of zero modes when compared to the original Dirac determinant, which does not contain  the factor $e^{i(\pi/2)(n_+-n_-)}$. However,  for the Grassman integral to give the product of the two Pfaffians, we need  reorder the $d[\chi]$ measure factors to get all the $d\chi_2$'s to the right of  the $d\chi_1$'s. For each mode number $n$ we have  
\be
d\bar \psi_{n,L}d\psi_{n,R}d\bar \psi_{n,R} d\psi_{n,L}= {\rm det}[{\mathcal C}]^{-1} d\chi_{n,1R}d\chi_{n,2R}d\chi_{n,2L}d\chi_{n,1L},
\ee
where the order of the factors in the measure on the LHS is mandated so that we get \hbox{${\rm Det}({\fsl D}+m)$}. The factors of ${\rm det}[{\mathcal C}]^{-1}$, one for each mode $n$, cancel the ``metric'' factor ${\rm Det} [C]= {\rm Det}[{\mathcal C}\otimes {\mathbb I}]$   that always occurs when we use eigenvalues of a linear  operator such as ${\fsl D}+m$ to compute the  Pfaffian of an associated  skew symmetric matrix such as ${\mathcal C}({\fsl D}+m)$ (Appendix \ref{SEC:pfaffian}).  
If all modes are present the remaining factors on the  RHS  can be  rearranged to get the Pfaffian without changing the sign. If there {\it is\/} a zero mode then 
  (taking into account that each zero mode occurs twice)  the factors of $(-1)$ that arise from each exchange of pairs of $d\chi$'s cancel the   extra sign from the $i\gamma^5$ mass. This resolves the apparent paradox that motivated the paper \cite{golterman}.

 \subsection{Euclidean-signature pseudo-Majorana fermions}
   If 
\be
{\fsl D} u_n= i\lambda_n u_n
\ee
then
\be
{\fsl D}^* u^*_n= -i\lambda_n u^*_n
\ee
or, assuming that any gauge fields are in real representations,
\be
{\fsl D} {\mathcal T}^{-1} u^*_n= -i\lambda_n {\mathcal T}^{-1}  u^*_n .
\ee
If $\lambda_n\ne 0$ we will have $u_n$ and $ {\mathcal T}^{-1}  u^*_n$ orthogonal because they have different eigenvalues. Let let us choose $u_n$ to be the positive-$\lambda_n$  eigenfunctions and consider the cases $d$=8, 9, 10 (mod 8) in which ${\mathcal T}$ is symmetric and we can consistently impose the Minkowski signature {\it pseudo-Majorana condition\/}
and take the action to be
\be
S= \frac 12 \int d^dx\, \psi^T {\mathcal T} {\fsl D} \psi .
\ee
Note that were  ${\mathcal T}$  {\it skew\/}-symmetric, then
\be
\frac 12 \int d^dx\, \psi^T {\mathcal T} {\fsl D} \psi  =0.
\ee
Consequently  Euclidean-signature pseudo-Majoranas are {\it only\/} available in the same dimensions as Minkowski-signature pseudo-Majoranas. 

If there are no zero modes we can   expand  
\bea
\psi (x)&=& \sum_n[\xi_n u_n(x) + \eta_n  ({\mathcal T}^{-1}  u^*_n(x))]\nonumber\\
\psi^T(x) &=& \sum_n[\xi_n u^T_n(x) + \eta_n  (  u^\dagger_n(x){\mathcal T}^{-1})]
 \eea
 and find that 
 \be
\frac 12 \int d^dx\, \psi^T {\mathcal T} {\fsl D} \psi  = \sum_{\lambda_n>0}( i\lambda_n) \eta_n \xi_n.
\ee
As in the  ordinary Majorana case, the partition function is the product of only half the eigenvalues, so  again we get a square-root of the full Dirac determinant which we expect to   identify with the Pfaffian ${\rm Pf}[{\mathcal T} {\fsl D}] $ of the skew-symmetric kernel ${\mathcal T} {\fsl D}$. 

What is different from  the ordinary Majorana case   is that  
we cannot add a mass term by taking  
\be
S=   \frac 12 \int d^dx\, \psi^T {\mathcal T} ({\fsl D}+m) \psi
\ee 
because $\psi^T {\mathcal T}\psi=0$ by the symmetry of $\mathcal T$. Consequently  Euclidean pseudo-Majorana fermions are necessarily massless --- just as are their Minkowski bretheren.  

As second consequence of $\psi^T {\mathcal T}\psi=0$ is that establishing  that the  essential unimodularity of the diagonalizing matrices  requires a slightly different tactic.  If we replace the anticommuting $\xi_n$ and $\eta_n$ by commuting variables $X_n$ and $Y_n$ and define 
 \bea
\phi_\alpha (x)&=& \sum_{\n}[X_n u_{\alpha n}(x) + Y_n  ({\mathcal T}^{-1}_{\alpha \beta}  u^*_{\beta n}(x))]\nonumber\\
\phi^T_\alpha (x) &=& \sum_n[X_n u^T_{\alpha n}(x) + \eta_n  (  u^\dagger_{\beta n}(x){\mathcal T}_{\beta\alpha}^{-1})].
 \eea 
Then
\be
\int d^dx \phi^T(x) {\mathcal T} \phi(x) = \sum_n (X_nY_n+Y_n X_n),
\ee
independently of particular form of the $u_n$ eigenfunctions. All diagonalizing transformations ${B}$   therefore preserve the same (non-positive definite) symmetric form and can differ only by factors drawn from   some  orthogonal group. Orthogonal matrices obey  ${\rm det}[B]^2=1$, so, unlike the ordinary Majorana case where the ${ B}$ matrices differ  by symplectic (and therefore unimodular) matrix factors, here we have the possibility ${\rm det}[B]=-1$ and the  Pfaffian changing sign. Indeed we have already seen an inherent  sign ambiguity because we arbitrarily assigned the positive eigenvalue $\lambda_n$  to $u_n$ and $\xi_n$, rather than to $ {\mathcal T}^{-1}  u^*_n$  and $\eta_n$. This ambiguity is a potential   source of global anomalies: if  we smoothly interpolate between two gauge-equivalent background fields (which necessarily have the same set of $\lambda_n$), and  an odd number of $\lambda_n$ change sign during the interpolation then the partition function changes sign  and   the  theory   inconsistent \cite{witten-path-integral}.

 \subsection{\bf Majorana-Weyl  fermions}

In $2\, ({\rm mod} \,8)$ dimensions  $\Gamma^5\equiv \gamma^{8k+3}$ obeys ${\mathcal C}\Gamma^5 {\mathcal C}^{-1} =-(\Gamma^5)^T$ and therefore  ${\mathcal C}{\fsl D}\Gamma^5= (\Gamma^5)^T{\mathcal C}{\fsl D}$. We can thus decompose 
\be
\psi^T {\mathcal C}( {\fsl D}+m)\psi=\psi^T_R {\mathcal C}( {\fsl D})\psi_R+\psi^T_L {\mathcal C}( {\fsl D})\psi_L+ m ( \psi_R^T{\mathcal C}\psi_L+ \psi_L^T{\mathcal C}\psi_R).
\ee
If $m=0$,  we  may retain only  one of the the right or left fields, in which case we  have a Euclidean {\it Majorana-Weyl} fermion.
 
 \subsection{ Rokhlin's theorem} 
Recall that in $d=4\, ({\rm mod\,}8)$ the  ${\mathcal T} $ matrix is antisymmetric and obeys ${\mathcal T}\Gamma^{5}  {\mathcal T}^{-1}= (\Gamma^5)^T= (\Gamma^5)^*$, where $\Gamma^5\equiv \gamma^{8k+5}$. Assume that no gauge fields are present ({\it i.e \/} gravity only), then  similar algebra to the previous section shows  that if $u_n$ obeys 
\be  
{\fsl D}u_n=i\lambda_n u_n,
\ee
then 
\be
{\fsl D} {\mathcal T}^{-1}u^*_n=-i\lambda_n {\mathcal T}^{-1} u^*_n. 
\ee
In particular, if $\lambda_n=0$ then ${\mathcal T}^{-1}u^*_n$ is also a zero mode, is orthogonal to 
$u_n$, and has the same $\gamma^{8k+5}$ eigenvalue. 
It follows that chiral zero modes come in pairs, and so the Dirac index 
\be
n_+-n_-=\int_M \hat A(R)
\ee
is an {\it even\/} integer.  
In dimension 4 this result implies Rokhlin's theorem that the signature of a 4-dimensional spin manifold is divisible by 16. This is  because when $d=4$ 
\bea
\hat A_1&=& -\frac 1{24} {\mathfrak p}_1, \,\quad\hbox{Dirac Index},  \nonumber\\
L_1&=& +\frac 13 {\mathfrak p}_1, \qquad\hbox{Signature},\nonumber
\eea
where 
\be
{\mathfrak p}_1=-\frac 1{(2\pi)^2}\tr\left\{\frac 12 {R} \wedge  {R}  \right\},
\ee 
is the four-form Pontryagin class. We see that the $\hat A$-genus whose integral gives $n_+-n_-$  evaluates to minus one-eighth of the signature.

Observe that  iterating the antilinear map $u_n\to {\mathcal T}^{-1} u_n^*$ twice gives
\be
u_n\to {\mathcal T}^{-1}[{\mathcal T}^{-1} u_n^*]^*= -u_n,
\ee
where we have used ${\mathcal T}^*=[{\mathcal T}^{-1}]^T= -{\mathcal T}^{-1}$. Thus our map gives rise to a {\it quaternionic structure\/} --- {\it i.e.\/}\ an antilinear map that squares to minus the identity ---  on the zero mode space. This is how Rokhlin's theorem is explained in the mathematics literature.

We could, of course, have deduced the doubling of the zero modes from the Majorana doubling given by the antisymmetric ${\mathcal C}$. This also gives rise to a quaternionic structure. Indeed in $d=4$ (mod 8)  we have 
\be
{\mathcal C}= \Gamma^5 {\mathcal T},
\ee
so we can take 
\be
Q_\pm \stackrel{\rm def}{=} \frac12 (1\pm \Gamma^5) {\mathcal T} 
\ee
to be  pair of independent quaternionic structures, one for each of the $\Gamma^5\to \pm 1$   subspaces of chiral zero-modes.

 \subsection{Symplectic Majorana fermions}
 
   In $d=$5, 6, 7 (mod 8) Euclidean dimensions  neither Majorana nor pseudo-Majorana  fermion actions  can be constructed. There is however the option of {\it  symplectic Majorana\/} fermions.  To obtain  these we start from  a {\it pair\/} of fermions $\psi_1$, $\psi_2$. In $d=7$ (mod 8) where $\mathcal C$ is symmetric we can set
\be
S= \frac 12 \int \psi_a^T {\mathcal C} \epsilon^{ab} ({\fsl D}+m)\psi_b d^{8k+7} x.
\ee
In  $d=5$ (mod 8) where $\mathcal T$ is antisymmetric we can take
\be
S= \frac 12 \int \psi_a^T {\mathcal T} \epsilon^{ab} {\fsl D}\psi_b d^{8k+5} x.
\ee
In $d=6$ (mod 8) we  can use either of ${\mathcal C}$ or ${\mathcal T}= {\mathcal C}\Gamma^5$, and if  the mass vanishes  we can have {\it symplectic Majorana-Weyl\/} fermions.

In all three cases the Minkowski-siganture Majorana constraint is 
\be
\psi_a= \epsilon_{ab} ({\mathcal C} \hbox{ or }{\mathcal T})^{-1}(\gamma^0)^T \psi_b^*.
\ee

\section{Multiplets, continuous symmetries, and  Bott periodicity}

We have so far considered  Majorana fields  one at a time. Let us add  a   ``flavour'' index taking values $n=1,\ldots, N$ to the $\psi$'s  and   consider  the possible symmetry groups that the resulting  $N$-tuplets can possess.

\subsection{Massless fields}

For massless fields, the resulting symmetry groups are      displayed in  table \ref{TAB:flavour}. 
\begin{table}
\begin{center}
\begin{tabular}{|c|c|c|c|c|}
\hline\hline
d (mod 8) & ${\mathcal T}$& ${\mathcal C}$&G & H  \\
\hline
0  &S &S&${\rm U}(N)$&${\rm Sp}(N/2)$\\
1 &S& &${\rm O}(N)$ &${\rm U}(N/2)$\\
2& S & A&${\rm O}(N)\times {\rm O}(N) $&${\rm O}(N)$\\
3& &A&${\rm O}(N)$&${\rm O}(N/2)\times {\rm O}(N/2)$\\
4 &A&A&$ {\rm U}(N)$& ${\rm O}(N)$\\
5 &A&&${\rm Sp}(N)$&${\rm U}(N)$\\
6 &A&S&$ {\rm Sp}(N)\times {\rm Sp}(N)$&${\rm Sp}(N)$\\
7& &S& ${\rm Sp}(N)$& ${\rm Sp}(N/2)\times {\rm Sp}(N/2)$\\
\hline\hline
\end{tabular}
\end{center}
\caption{\sl The Minkowski signature flavour groups $G$ that preserve the action of massless Majorana fermions in $d$ spacetime dimensions.  In the table   ${\rm Sp}(N) \equiv  {\rm Sp}(2N, {\mathbb C})\cap {\rm U}(2N)$ is the unitary symplectic group. The subgroups $H$ are those that  survive the addition of suitable mass terms.} 
\label{TAB:flavour}
\end{table}

Not all of these groups are immediately obvious. Consider for example $d=4$ (mod 8). The Euclidean Majorana action is invariant under separate chiral transformation $\psi_R\to U \psi_R$, $\psi_L =V\psi_L$  provided the   $N$-by-$N$ matrices $U$, $V$, acting on the new indices obey $U^T V={\mathbb I}_N$.  In Euclidean signature there is no obvious reason  to impose any reality conditions \cite{wetterich}  on $U$ and $V$, but  when   we desire  the invariance  to extend to a symmetry  in Minkowski signature where $\psi = {\mathcal C}^{-1} (\gamma^0)^T \psi^*$ preserving this relation   requires that $V=U^*$. This condition makes  $U^\dagger U={\mathbb I}_N$  and hence $U\in {\rm U}(N)$. 

In $d=5,6,7$ preserving the kinetic term requires $\psi \to V\psi$ with $V$ a $2N$-by-$2N$ matrix obeying  $V^T \Omega V=\Omega$. Here 
$\Omega$ is block diagonal with $N$ copies on the diagonal of the two-by-two matrix with entries $\epsilon_{ab}$. This condition requires $V$ to be in ${\rm Sp}(2N, {\mathcal C})$,  and, as $\Omega^{-1}=-\Omega$, can also be written as 
\be
-\Omega V\Omega = (V^T)^{-1}.
\label{EQ:symplectic1}
\ee 
The Minkowski-signature reality conditions are of the form  $\psi =\Omega \psi^*$ (ignoring the spinor indices) and  preserving them   requires   $-\Omega V \Omega=V^*$. Consistency with  equation (\ref{EQ:symplectic1})  now needs   $(V^T)^{-1}=V^*$, meaning that   $V$  is  a  unitary matrix.   $V$ is therefore an element of the unitary symplectic group ${\rm Sp}(N) \equiv  {\rm Sp}(2N, {\mathbb C})\cap {\rm U}(2N)$.

The other cases are more straightforward except that in 2, 6 (mod 8) dimensions the products ${\rm O}(N)\times {\rm O}(N) $ and  $ {\rm Sp}(N)\times {\rm Sp}(N)$  arise as   a consequence of the existence of Majorana-Weyl fermions in those dimensions.

We can    elevate the global flavour symmetries to local  symmetries in which case  the flavour indices become ``colour" indices  that couple to gauge fields in the fundamental representation of the corresponding group. It may seem unlikely that a four-dimensional Majorana field can couple to a ${\rm U}(N)$ gauge field but, because the left- and right-handed  fermions transform under conjugate representations, the induced coupling is to  an {\it axial\/}  gauge field. Unlike the  vector current $\bar\psi {\mathcal C}\lambda_a\gamma^\mu \psi$,  the axial current $\bar\psi {\mathcal C}\lambda_a\gamma^5\gamma^\mu \psi$  is not identically zero. An example of such a Majorana ${\rm U}(1)$ axial gauge-field  coupling appears  in models of   topological  superconductors \cite{qi,stone-lopes,hansson}.

 What is most interesting about the  pattern of groups in  table \ref{TAB:flavour}  is that,  starting arbitrarily  from $d=3$ (mod 8) and  with suitable $d$-dependent choices of $N$, we have a natural sequence of group embeddings 
\bea
\ldots {\rm O}(16N)\supset {\rm U}(8N)\supset {\rm Sp}(4N)\supset  {\rm Sp}(2N)\times {\rm Sp}(2N)\supset {\rm Sp}(2N)\nonumber\\ \supset {\rm U}(2N)
\supset {\rm O}(2N) \supset {\rm O}(N)\times {\rm O}(N)\supset {\rm O}(N)\ldots \nonumber
\eea
of each group in its predecessor. 
Starting from $d=3$ is not  essential. The  sequence  can be extended to the left, and to the right when $N$ is a suitably large  power of two. The pattern repeats with period eight, and the  successive quotients of the groups are the reductive symmetric spaces  that appear in  Bott's  periodicity theorem for stable homotopy groups.This theorem is  displayed in  table \ref{TAB:bott}.  The same pattern appears, and for the same reason --- the trading  of spinor indices  for  flavour indices  --- in the sequence of ``R" symmetries  that appear during  the  reduction of $N=1$ supersymmetry algebras to larger $N$ algebras in lower dimensions \cite{proeyen,weinbergQ}. 

\begin{table}  
\begin{center}
\begin{tabular}{|c|c|c|c|c|}
\hline\hline
Cartan Label&Coset $G/H$&$\pi_0(G/H)$ \\
\hline
A1  & ${\rm U}(N)/{\rm O}(N)$&$0$\\
BD1 &${\rm O}(2N)/{\rm O}(N)\times {\rm O}(N)$&${\mathbb Z}$ \\
D& ${\rm O}(2N)\times {\rm O}(2N)/{\rm O}(2N)\simeq {\rm O}(2N)$& ${\mathbb Z}_2$\\
DIII& $ {\rm O}(2N)/{\rm U}(N)$& ${\mathbb Z}_2$\\
AII &  $ {\rm U}(2N)/{\rm Sp}(N)$&$0$\\
CII & ${\rm Sp}(2N)/{\rm Sp}(N)\times {\rm Sp}(N)$&${\mathbb Z}$\\
C & ${\rm Sp}(N)\times {\rm Sp}(N)/{\rm Sp}(N)\simeq {\rm Sp}(N)$& 0\\
CI&  ${\rm Sp}(N)/{\rm U}(N)$& 0\\
\hline\hline
\end{tabular}
\end{center}
\caption{\sl The Cartan classification of the symmetric spaces arising from  quotienting  each of the  flavour symmetry groups by its successor starting from $d=0$ (mod 8). The Bott periodicity theorem asserts  that, for sufficiently  large $N$,  the homotopy group $\pi_{k}(G/H)$ is isomorphic to  $\pi_0(G/H)$ of the symmetric  space $k$ rows (mod 8)  lower down   the table.} 
\label{TAB:bott}
\end{table}

 \subsection{Mass terms and symmetry breaking}

In 4 (mod 8) dimensions the addition of a $m \psi^T{\mathcal C}\psi$  mass  term to the Majorana action breaks the chiral  flavour symmetry from ${\rm U}(N)$ down to ${\rm O}(N)$. This is  because preserving such a mass term requires the matrices $U$ and $V=U^*$ of the previous section to be equal. Note  that  ${\rm O}(N)$ is the group in the Bott cycle that {\it precedes\/} the ${\rm U}(N)$ in row 4 of table \ref{TAB:flavour}.  This backstepping  is part of a general pattern.  

To illustrate this pattern   consider some  possible symmetry breaking mass terms\footnote{The first line of this list shows that  prohibition of a mass for pseudo-Majorana fermions can   be evaded   when  there is more than one of them.}:

\bea
\psi^T{\mathcal T}M \psi, \quad &&\hbox{$M$ skew-symmetric in $d=0,1$ (mod 8) dimensions,}\nonumber\\
\psi^T{\mathcal C} M\psi, \quad &&\hbox{$M$  symmetric in $d=3,4$ (mod 8) dimensions}\nonumber\\
\psi^T{\mathcal T}\Omega M \psi, \quad &&\hbox{$\Omega M$ symmetric in $d=5$ (mod 8) dimensions,}\nonumber\\
\psi^T{\mathcal C}\Omega M \psi, \quad &&\hbox{$\Omega M$ skew-symmetric in $d=7$ (mod 8) dimensions.}
\nonumber
\eea

For a moment let us assume the maximally symmetric situation where  $M= {\rm diag}(m,\ldots m) $ in $d=3,4$  (mod 8) and $M= \Omega \, {\rm diag}(m,\ldots m)  $ in $d= 5$ (mod 8) dimensions.

In $d = 5$ (mod 8) dimensions  
preserving the mass term $\psi^T{\mathcal T} {\rm diag}(m,\ldots m) \psi $ requires the unitary symplectic matrix $V$  to obey $V^TV={\mathbb I}$ {\it i.e.\/}\ to be an orthogonal matrix. Consequently  the  ${\rm Sp}(N)
\simeq {\rm Sp}(2N,{\mathbb C})\cap {\rm U}(2N)$ flavour  group is broken to  ${\rm Sp}(2N,{\mathbb R})\cap {\rm O}(2N)\simeq {\rm U}(N)$.  

Similarly, in $d= 0$ (mod 8) dimensions  $M$ must be  skew, and if  non-degenerate,  $N$ must be an even number. Preserving the skew symmetric form  $\psi^T{\mathcal T}M \psi $ requires a symplectic group.  The symmetry is  therefore  reduced from  ${\rm U}(2N)$ to its subgroup
${\rm Sp}(2N,{\mathbb C})\cap {\rm U}(2N)\simeq {\rm Sp}(N)$.  In both these  cases the preserved sub group $H\subseteq G$ is  again the preceding one  the Bott cycle.  In $d=3,7$ (mod 8) we have an {\it option} of taking $N_-$ of the of  diagonal elements of the symmetric matrix to be $-m$ and the remaining $N_+=N-N_-$ to be be $+m$, so
so in d=3 (mod 8) we can take $O(N)\to O(N_+)\times O(N_-)$, and similarly the $d=7$ (mod 8) symplectic  case. 

 If the original symmetry group is of the form $G_{\rm gauge}\otimes  G_{\rm flavour}$ and the mass terms arise from spontaneous breaking of the flavour symmetry $G_{\rm flavour} \to H_{\rm flavour}$ by the gauge interactions,  the orbit of equivalent vacua   is the appropriate symmetric space $G_{\rm flavour}/H_{\rm flavour}$. 

\section{Discrete symmetries} 
\label{SEC:discrete} 

\subsection{Intrinsic parity of  Dirac and Majorana fermions}

  In even space-time dimensions parity is defined by ${\textsf P}:(t,{\bf x}) \mapsto (t,-{\bf x})$. In  the mostly-minus metric ${\textsf P}$  is implemented on spinor-valued fields  as 
\be
 {\textsf P}: \psi(t, {\bf x})\mapsto \psi^p(t,{\bf x})= \eta \gamma^0 \psi(t,-{\bf x}),
\ee
where the phase $\eta$ is the particle's {\it intrinsic parity\/}. We usually take $\eta=\pm 1$ so that that  ${\textsf P}^2={\rm id}$. 
However if ${\textsf  P}$ is to be compatible with the Majorana condition $\psi^c=\psi$ and if we require  $(\psi^c)^p=(\psi^p)^c$ then we must have  same parity transformation rule for $\psi$ and $\psi^c$. Let us see what this requires.

 Using $(\psi^p(t,{\bf x}))^*= \eta^* (\gamma^0)^* \psi^*(t,-{\bf x})$ we have
\be
\eta\gamma^0 [{\mathcal C}^{-1} (\gamma^0)^T\psi^*(t,-{\bf x})]= {\mathcal C}^{-1}(\gamma^0)^T\eta^* (\gamma^0)^* \psi^*(t,-{\bf x})
\ee
which reduces to 
\be
\eta {\mathcal C}\gamma^0 {\mathcal C}^{-1} (\gamma^0)^T = \eta^* (\gamma^0)^T (\gamma^0)^*,
\ee
or
\be
-\eta  (\gamma^0)^T = \eta^*  (\gamma^0)^*.
\ee
Since $\gamma^0$ is Hermitian  in the mostly minus metric we see that $\eta^*=-\eta$, and so for a Majorana particle we must have $\eta=\pm i$ and so  ${\textsf P}^2=-1$. 

If we have the freedom to allow 
\be
(\psi^c)^p(t,{\bf x}) = \eta^c \gamma^0 \psi^c(t,-{\bf x})
\ee
then the same algebra shows that   $\eta^c = -\eta^*$. For  particles that are distinct from their antiparticles  we are therefore allowed to have    $\eta$ to be $\pm 1$, but then   the parity of an antiparticle is  minus that of the particle. 

In the mostly-plus metric  parity is usually implemented by\footnote{Steven Weinberg's {\it The Quantum Theory of Fields\/}  \cite{weinberg} is the only text that I know that uses the mostly plus convention, and he has this ``$i$'' factor. Mark Srednicki's {\it Quantum Field Theory\/} \cite{sredniki} {claims\/} to use the  mostly-plus convention, but he  defines his Clifford algebra by $\gamma^\mu\gamma^\nu+\gamma^\nu\gamma^\mu=-2g^{\mu\nu}$, so his are   the mostly-minus gamma matrices.} 
\be
{\textsf P}:\psi(t, {\bf x})\mapsto \psi^p(t,{\bf x})=i  \eta \gamma^0 \psi(t,-{\bf x})  
\ee
The reason for the extra factor of $i$ is that when $\eta=1$ we again   want ${\textsf P}^2={\rm id}$, and the extra $i$ compensates for $(\gamma^0)^2=-1$.  For Majorana fields   we still  find that $\eta^*=-\eta$. 

\subsection { \textsf R symmetry}

In odd  space-time   dimension $d=2k+1$, the standard  parity operation $(t,{\bf x})\mapsto (t,-{\bf x})$  is an ${\rm SO}(2k)$  rotation, so ``parity''  is instead defined as the   inversion of  an odd number of the spatial coordinates.  In the case that we flip  only one direction Witten calls it \textsf{R} symmetry \cite{witten-path-integral}   
Let us define $\textsf R$ to invert  $x^1$ so 
\be
\textsf R:(t,x_1,x_2,\ldots x_{2k})\mapsto (t,-x_1,x_2,\ldots x_{2k})\equiv (t,\tilde {\bf x}).
\ee
In Euclidean signature the natural way to flip  the sign of  $\gamma^1$ only is by the using the  Clifford algebra ``twisted map" reflection in the plane perpendicular to the $x^1$ axis:
\be
\textsf R:\gamma^\mu \mapsto (-\gamma^1)\gamma^\mu \gamma^1   =\tilde  \gamma^\mu
\ee
To get 
\be
\textsf R:\bar\psi({\bf x}) \gamma^\mu \psi({\bf x}) \mapsto \bar\psi(\tilde {\bf x}) \tilde \gamma^\mu \psi(\tilde {\bf x})
\ee
we must therefore set
\bea
\textsf R:\psi({\bf x})&\mapsto& \gamma^1 \psi(\tilde{\bf x})\nonumber\\ 
\textsf R: \bar\psi({\bf x})&\mapsto& \bar\psi(\tilde{\bf x})(-\gamma^1).
 \eea
In Euclidean signature the  ${\textsf R}:  u_n({\bf x}) \mapsto \gamma^1 u_n(\tilde{\bf x})$ anticommutes with ${\fsl D}$ and so changes the sign of the corresponding eigenvalue $\lambda_n$

In Minkowski space  
consider  first the  mostly plus  metric in which $(\gamma^1)^2=1$ and $\gamma^1$ is Hermitian.  When   
\be
 \psi(t, {\bf x})\mapsto  \eta\gamma^1 \psi(t,\tilde{\bf x}) 
 \ee
 we have  
 \be
\bar\psi(t, {\bf x})\mapsto \overline{ \eta \gamma^1 \psi(t, \tilde {\bf x})} = \eta^* \psi^\dagger(t,\tilde {\bf x})(\gamma^1)^\dagger \gamma^0= \eta^* \bar\psi (t,\tilde {\bf x})(- \gamma^1)
\ee
and so  we have
\be
\textsf R: \bar\psi \gamma^\mu \psi \mapsto  \bar\psi \tilde \gamma^\mu \psi.
\ee

In the mostly-minus metric when  
$\textsf R$ acts on the Fermi fields as 
\be
 \psi(t,{\bf x})\mapsto  \eta \gamma^1\psi(t,\tilde {\bf x})
 \ee
 then 
 \be
\bar\psi(t, {\bf x})\mapsto \overline{ \eta \gamma^1 \psi(t, \tilde {\bf x})} = \eta^* \psi^*(t,\tilde {\bf x})(\gamma^1)^\dagger \gamma^0= \eta^* \bar\psi (t,\tilde {\bf x}) \gamma^1.
\ee
This appears to differ from the Euclidian ${\textsf R}$, but $(\gamma^1)^2=-1$, so we still have 
\be
\textsf R: \bar\psi \gamma^\mu \psi \mapsto  \bar\psi \tilde \gamma^\mu \psi.
\ee

As  $\partial_{\bf x}\mapsto \partial_{\tilde{\bf x}}$  both signature versions  of $\textsf R$ leave the kinetic part of the Dirac action invariant. 
However  
\be
{ \textsf R}:\bar\psi\psi \,\,{\mapsto} - \bar\psi\psi.
\ee
so the mass term is {\it not\/}   invariant. In even space-time dimensions we can undo the flip with a $\Gamma^5$ and so obtain the usual parity operation which leaves $m$ fixed. This option is not available  in odd space-time dimensions, where a mass is unavoidably parity-violating.

Requiring  that reflection commutes with charge conjugation  leads to $\eta_c = - \eta^*$. To see this compare
\be
[\psi^c(t,{\bf x})]^r= \eta_c \gamma^1 [{\mathcal C}^{-1} \gamma^{0T} \psi^*(t,\tilde {\bf x})]
\ee
with
\bea
[\psi^r(t,{\bf x})]^c &=& {\mathcal C}^{-1} \gamma^{0T} [\eta^*\gamma^{1*} \psi^*(t,\tilde {\bf x})]\nonumber\\
&=& -{\mathcal C} {\mathcal C} \gamma^0 {\mathcal C}^{-1} \gamma^{1*} \psi^*(t, \tilde {\bf x})\nonumber\\
&=& -\eta^* \gamma^0 {\mathcal C}^{-1} \gamma^{1*}\psi^*(t, \tilde {\bf x})\nonumber\\
&=& + \eta^* \gamma^0 {\mathcal C}^{-1} \gamma^{1T}\psi^*(t, \tilde {\bf x})\nonumber\\
&=& -\eta^* \gamma^0 {\mathcal C}^{-1} {\mathcal C} \gamma^{1}{\mathcal C}^{-1}\psi^*(t, \tilde {\bf x})\nonumber\\
&=& -\eta^* \gamma^0 \gamma^{1}{\mathcal C}^{-1}\psi^*(t, \tilde {\bf x})\nonumber\\
&=& +\eta^* \gamma^1 \gamma^{0}{\mathcal C}^{-1}\psi^*(t, \tilde {\bf x})\nonumber\\
&=& - \eta^* \gamma^1[{\mathcal C}^{-1} \gamma^{0T} \psi^*(t, \tilde {\bf x})].
\eea


\subsection{ Time reversal}

 At first sight time reversal should simply be an $\textsf R$ map applied to $x^0\equiv t$ rather than $x^1$. However such an $\textsf R$ reverses the direction  of particle trajectories in time and so converts particles into antiparticles.  The conventional  particle-physics (Wigner) time reversal operation  does   not charge-conjugate  and  so $\textsf T$ is defined by composing  an $\textsf R$ with a compensating charge conjugation operation. 
There is still a problem: time reversal  does not play nicely with the passage from Euclidean to Minkowski signature because,
as in non-relativistic quantum mechanics, time reversal must be  implemented on the many-particle Hilbert space by an {anti}unitary operator ${\mathfrak I}$.    

An operator $\Omega$ is said to be antiunitary with respect to a conjugate-symmetric sesquilinear inner product $\brak{\phantom -}{\phantom-}$ if  
\be
\brak{\Omega {\bf a}} {\Omega {\bf b}} =\brak{{\bf a}}{{\bf b}}^*= \brak{\bf b}{\bf a}. 
\ee
 Consider the vector
\be
{\bf X}= \Omega( \alpha {\bf a}+\beta {\bf b})- \alpha^* (\Omega {\bf a}) -\beta^* (\Omega{\bf b}).
\ee
Using   the definition of antiunitarity and the antilinearity of $\brak{\phantom -}{\phantom-}$ in its  first slot and linearity in the second,  we can expand out   $\|{\bf X}\|^2=\brak{{\bf X}}{{\bf X}}$ and find that it is zero. For a  positive definite  inner product a vanishing norm  implies  that ${\bf X}=0$, and so for such a product we have 
\be
\Omega( \alpha {\bf a}+\beta {\bf b})= \alpha^* (\Omega {\bf a}) +\beta^* (\Omega{\bf b}).
\ee
Thus an  antiunitary operator acting on a positive-definite Hilbert space is necessarily antilinear. 

One  consequence of the antilinearity  is that there is no way to define an  adjoint $\Omega^\dagger$.   The standard definition $\brak{\Omega^\dagger {\bf a}}{{\bf b}}= \brak{{\bf a}}{\Omega{\bf b}}$ leads to   
\be
\brak{{\bf b}}{{\bf a}}=\brak{\Omega  {\bf a}} {\Omega {\bf b}} \stackrel{?}{=} \brak{\Omega^\dagger \Omega {\bf a}}  {{\bf b}}
\ee
and  a contradiction:~the leftmost expression is antilinear in ${\bf b}$ while the rightmost is linear in ${\bf b}$.   A similar issue leads to 
\be
(\bra{\bf a} \Omega)\ket{\bf b}\ne \bra{\bf a}(\Omega \ket{\bf b})
\ee
and so makes Dirac notation ``matrix elements'' $\eval{{\bf a}}{\Omega}{{\bf b}}$ ambiguous. It also   prevents us from defining  a  left action  of  $\Omega$ on bra vectors $\bra{\bf a}$.
Instead we have the useful identity\footnote{Some sources---for example the Wikipedia article on antiunitary operators---define an ``$\Omega^\dagger$" by equating  it to $\Omega^{-1}$. I think that this notation is dangerously   confusing.}
\be
\brak{{\bf b}}{\Omega{\bf a}}= \brak{{\bf a}}{\Omega^{-1} {\bf b}}.
\ee

Another useful result is that if  $A$ is a linear operator then so is $\Omega^{-1}A\Omega$, and we can compute its adjoint as follows
\be
\brak{{\bf b}}{(\Omega^{-1}A\Omega) {\bf a}} =\brak{\Omega {\bf b}}{A\Omega{\bf a} }^*= \brak{A^\dagger \Omega {\bf b}}{\Omega \bf a}^*=\brak{(\Omega^{-1}A^\dagger \Omega){\bf b}}{\bf a} 
\ee
so
\be
(\Omega^{-1}A\Omega)^\dagger =\Omega^{-1}A^\dagger\Omega.
\ee

In  the mostly minus Minkowski metric the   time reversal operator ${\mathfrak I}$  is usually taken to  acts on Dirac  field operators  as 
\bea
{\mathfrak I}^{-1}  \psi({\bf x},t) {\mathfrak I} &=& \eta_T {\mathcal T}\psi({\bf x},-t)\nonumber\\ 
{\mathfrak I}^{-1}{\bar \psi}({\bf x},t) {\mathfrak I}&=& \eta^*_T {\bar\psi}({\bf x},-t){\mathcal T}^{-1},
\eea
where $\eta_T$ is a phase.  Despite the antilinearity of ${\mathfrak I}$ the field operator is not Hermitian-conjugated: time reversal  changes the sign of  momentum and the spin, but does  {\it not\/} change particle to antiparticle. This, however, is the action on the field {\it operator\/}. The action  of ${\mathfrak I}$ on wavefunctions {\it does\/} involve complex conjugation, and will be described later.

We can decompose the action of ${\mathfrak I}$ into a composition  of $\textsf T=\textsf C\textsf  R$ followed by  complex conjugation:
\bea
\psi({\bf x},t)&\stackrel{\textsf R}{\mapsto}& \gamma^0 \psi({\bf x},-t)\nonumber\\
 \gamma^0 \psi({\bf x},-t) &\stackrel{\textsf C}{\mapsto}& \eta {\mathcal T}^{-1}(\gamma^0)^T (\gamma^{0} )^*\psi^*({\bf x},-t)=\eta {\mathcal T}^{-1}\psi^*({\bf x},-t) \nonumber\\
\eta {\mathcal T}^{-1}\psi^*({\bf x},-t)  &\stackrel{*}{\mapsto} &\eta^*({\mathcal T}^{-1})^* \psi({\bf x},-t)\nonumber\\
 &=& 
 \lambda \eta^* {\mathcal T} \psi({\bf x},-t),
 \eea
 where ${\mathcal T}^T=\lambda {\mathcal T}$.
We have elected  to  use  the ${\mathcal T}$ version of charge conjugation rather than the ${\mathcal  C}$ version because a $\mathcal T$ conjugation inverts $\bar\psi\psi$ and so undoes the $\bar\psi\psi$ inversion due to the ${\textsf R}$. As a result 
\be
\bar\psi({\bf  x},t)\psi({\bf x},t) \mapsto {\mathfrak I}^{-1}\bar \psi({\bf x},t)\psi ({\bf x},t) {\mathfrak I}= {\mathfrak I}\bar  \psi({\bf x},t){\mathfrak I} ^{-1}{\mathfrak I}^{-1} \psi({\bf x},t){\mathfrak I}= \bar\psi({\bf x},-t)\psi({\bf x},-t).
\ee
A $\Gamma^5$ chiral mass term {\it does\/} change sign.

The transformation of $\bar \psi$ follows that of $\psi$ {\it via\/}
\be
({\mathfrak I}^{-1}A{\mathfrak I} )^* ={\mathfrak I}^{-1}A^*{\mathfrak I}.
\ee
We  use  $*$ instead of $\dagger$ to indicate that the  Hermitian adjoint in the quantum-state Hilbert space does not transpose  column-matrix spinors to  row-matrix spinors.
Then
\be
{\mathfrak I^{-1}}  \psi({\bf x},t) {\mathfrak I}= \eta_T {\mathcal T}\psi({\bf x},-t)\Rightarrow 
{\mathfrak I}^{-1}  \psi^*({\bf x},t) {\mathfrak I} = \eta^*_T {\mathcal T}^*\psi^*({\bf x},-t).
\ee
Transposing and using antilinearity
\bea
{\mathfrak I}^{-1}  \psi^\dagger({\bf x},t)\gamma^0 {\mathfrak I} &=& \eta^*_T \psi^\dagger({\bf x},-t) {\mathcal T}^\dagger(\gamma^0)^*\nonumber\\
&=& \eta^*_T \bar \psi({\bf x},-t)\gamma^0 {\mathcal T}^{-1}(\gamma^0)^*\nonumber\\
&=& \eta^*_T \bar \psi({\bf x},-t){\mathcal T}^{-1}{\mathcal T}\gamma^0 {\mathcal T}^{-1}(\gamma^0)^*\nonumber\\
&=& \eta^*_T \bar \psi({\bf x},-t){\mathcal T}^{-1} (\gamma^0)^T(\gamma^0)^*\nonumber\\
&=& \eta^*_T \bar \psi({\bf x},-t){\mathcal T}^{-1}.
\eea

The time reversal of the current is 
\bea
{\mathfrak I}^{-1}\bar\psi({\bf x},t) \gamma^\mu \psi({\bf x},t){\mathfrak I}&=&  {\mathfrak I}^{-1}\bar\psi ({\bf x},t){\mathfrak I}{\mathfrak I}^{-1} \gamma^\mu \psi({\bf x},t){\mathfrak I}\nonumber\\
&=& {\mathfrak I}^{-1}\bar\psi {\mathfrak I}(\gamma^\mu)^*{\mathfrak I}^{-1}\psi({\bf x},t){\mathfrak I}\nonumber\\
&=& \bar \psi({\bf x},-t) {\mathcal T}^{-1} (\gamma^\mu)^* {\mathcal T}\psi({\bf x},-t) \nonumber\\
&=& \bar \psi({\bf x},-t) {\mathcal T}^{-1} (\gamma^{\mu\dagger})^T {\mathcal T}\psi({\bf x},-t)\nonumber\\
&=&\pm \bar \psi({\bf x},-t) {\mathcal T}^{-1} (\gamma^{\mu})^T {\mathcal T}\psi({\bf x},-t)\nonumber\\
&=& \pm \bar \psi({\bf x},-t) \gamma^{\mu} \psi({\bf x},-t).
\eea
Here $\gamma^{\mu\dagger}= \pm\gamma^\mu$ so we have   $+$ for the Hermitian $\gamma^0$, so the charge is not altered,  and $-1$ for the antiHermitian  $\gamma^a$ which changes the sign of the  the spatial  current.

If we act   twice we find 
\be
 {\mathfrak I}^{-2}\psi({\bf x},t) {\mathfrak I}^2= |\eta_T|^2 {\mathcal T} {\mathcal T}^* \psi({\bf x},t).
\ee
Now ${\mathcal T}^{-1}= {\mathcal T}^\dagger = ({\mathcal T}^T)^* = \lambda {\mathcal T}^*$ so
\be
 {\mathfrak I}^{-2}\psi({\bf x},t) {\mathfrak I}^2 = \lambda \psi({\bf x},t).
\ee
Thus conjugating  by ${\mathfrak I}$ twice gives a $-1$ in 4, 5, 6 (mod 8) dimensions in which ${\mathcal T}$ is antisymmetric. As the vacuum is left fixed 
\be
{\mathfrak I}\ket{0}= \ket{0}
\ee
and  the field operators change the fermion number by $\pm 1$, the    $-1$  means that when ${\mathfrak I}$ acts  on the many-particle Hilbert space  we have  ${\mathfrak I}^2 =(-1)^F\,{\rm id}$ where $F$ is the fermion number. We get  a $+1$, and hence ${\mathfrak I}^2= {\rm id}$,  in  0, 1, 2 (mod 8) dimensions in which ${\mathcal T}$ is symmetric.  

In 3 and 7 (mod 8) dimensions the  ${\mathcal T}$ matrix does not exist.  We can however use the ${\mathcal C}$ charge-conjugation operation to define an alternative  time reversal 
\bea
{\mathfrak J} ^{-1} \psi({\bf x},t) {\mathfrak J} &=& \phantom- \eta_T {\mathcal C}\psi({\bf x},-t)\nonumber\\ 
{\mathfrak J}^{-1}{\bar \psi}({\bf x},t) {\mathfrak J} &=& -\eta^*_T {\bar\psi}({\bf x},-t){\mathcal C}^{-1},
\eea
at the expense of flipping the  the sign of $\bar\psi\psi$. Thus a mass term is necessarily time-reversal-symmetry violating in  3 and 7 (mod 8) dimensions.  Acting twice, this time reversal gives a $(-1)^F$ in 3 (mod 8) dimensions and a plus sign in 7 (mod 8). 

\subsubsection{${\textsf T}$ and anomaly inflow}

 Consider  a chiral (Weyl) fermion in space-time dimension $d=2k$ and interacting with an abelian gauge field $A_\mu$.  This is an anomalous theory  in which the anomaly can be be accounted for by  current flowing  into the $d=2k$   dimensional    surface from the  $D=2k+1$ bulk at a rate \cite{callan-harvey} 
\be
J^{2k+1}=  \frac 1{(2\pi)^k k!}  {\rm sgn}(M)\epsilon^{2k+1,i_1,\ldots, i_{2k}} F_{i_1i_2}\cdots F_{i_{2k-1}i_{2k}}.
\label{EQ:inflow}
\ee
Here  $M$ is a large Dirac mass in  the $2k+1$ dimensional theory. If we reverse time, the direction of this flow should  reverse.  How does this reversal relate to our discussion so far? 

In Minkowski signature  time reversal acts on the components of the gauge field as   
\be
A_0\mapsto A_0,\quad{\bf A}\mapsto -{\bf A}.
\ee
From this we see that  the ``electric field'' 
\be
F_{0i}= \partial_0 A_i - \partial_i A_0
\ee 
is unaffected by  time reversal, but all other (magnetic) $F^{ij}$ change sign. Consequently  the gauge-field $2k$-form $F^k$ in equation (\ref{EQ:inflow}) is time-reversal invariant in  $2k+1=$  3 and 7 (mod 8) bulk dimensions and changes sign in 1 and 5 dimensions (mod 8). The reversal of $J^{2k+1}$ is therefore accounted for  by the mass $M$ changing sign in 3 and 7 (mod 8) and by the $F^k$ factor changing sign in the other odd dimensions.   

A  change in sign of the $2k+1$ bulk theory Dirac mass should also cause a change in the $\Gamma^5$ chirality of the surface-trapped $2k$-dimensional  fermions. An inspection of the  table shows that it is precisely in 2 and 6 dimensions that $\Gamma^5$ anticommutes with both ${\mathcal C}$  and ${\mathcal  T}$, and so the change in chirality is consistent with the   
\be
\psi(t,{\bf x})\mapsto ({\mathcal C}  \hbox { or } {\mathcal  T}) \psi(-t,{\bf x})
\ee
time reversal transformation.

\subsubsection {The action of $\textsf T$ on wavefunctions}

 The $c$-number spinor wavefunction corresponding to a single particle state $\ket{\phi}$ is 
\be
\phi({\bf x},t)= \eval{0}{\psi({\bf x},t)}{\phi}.
\ee
The wavefunction of the time reversed state is then
\bea
\bra{0} ({\psi({\bf x},t)}{\mathfrak I}\ket{\phi}) &=& 
 \eval{0}{({\mathfrak I}^{-2}{\psi({\bf x},t)}{\mathfrak I}} {\phi})\nonumber\\ 
 &=&   \eval{0}{{\mathfrak I}^{-1}({\mathfrak I}^{-1}{\psi({\bf x},t)}{\mathfrak I}} {\phi})\nonumber\\
 &=& (\eval {0}{{\mathfrak I}^{-1}\psi({\bf x},t){\mathfrak I}}{ \phi})^*\nonumber\\
&=&(\eval{0}{\eta_T {\mathcal T}\psi({\bf x},-t)}{ \phi})^*\nonumber\\
&=&  \eta^*_T {\mathcal T}^* (\eval{0}{\psi({\bf x},-t)}{ \phi})^*\nonumber\\
&=&\lambda \eta^*_T {\mathcal T}^{-1}\phi^*({\bf x},-t).
\eea
In the first line we used ${\mathfrak I}^{2}=(\lambda)^F$ with fermion number $F=0$, and  in passing from the second  to third  line we  used 
\be
\eval{0}{{\mathfrak I}^{-1}}{{\bf v}}\equiv \eval{0}{({\mathfrak I}^{-1}}{{\bf v}})= \eval{{\bf v}}{({\mathfrak I}} {0})=\dbrak{{\bf v}}{0}= \dbrak{0}{{\bf v}}^*. 
\ee
The result is that, in contrast with the transformation of the operator, the single-particle wavefunction {\it is\/} complex-conjugated. 

\section{Conclusion}  We have shown that  for  each class of   Majorana fermions   (original, pseudo and symplectic)  there is a Euclidean-signature  Grassmann action integral that exists in precisely the same dimensions as the corresponding  Minkowski-signature Majorana fermion fields . We have described how the pattern  of dimensions   in which these classes exist, and the possible continuous symmetries that they might possess, is controlled by the same   eightfold  Bott periodicity that appears in many areas of mathematics and physics. We have also related the manner in which the discrete C, P, T,    manifest themselves in both signatures to this   periodicity.

\section{Acknowledgements} This work was not directly supported by any funding agency, but it would not have been possible without resources provided by the Department of Physics at the University of Illinois at Urbana-Champaign.

\appendix
\appendixpage

\section{Berezin Integrals}
\label{SEC:berezin}

The path integral for    fermions  requires  a formal  integration over  Grassmann-valued fields.   Felix Berezin's  recipe   for  this process is  purely algebraic but  is called ``integration'' because its   output  mirrors, up to signs, the result of the corresponding analytic operation on real and complex variables.  The  general   Grassmann/Berezin integral   requires the sophisticated mathematics of sheaf theory \cite{zirnbauer3}, but  we require  only   ``Gaussian"   integrals, and these  are relatively straightforward.

\subsection{Finite number of variables}
\label{SEC:finite}

 If  $\bar \psi_\alpha$ and     $\psi^\beta$, $\alpha,\beta=1,\ldots, N$ are a  set of anticommuting Grassmann variables, we  define their    Berezin  integral  by setting   
 \be
  \int [d\bar\psi d\psi] \,  \bar\psi_1 \psi^1 \cdots  \bar\psi_N \psi^N\equiv \int \left[\prod_{\alpha=1}^N d\bar \psi_\alpha  d \psi^\alpha\right] \bar\psi_1 \psi^1 \cdots  \bar\psi_N \psi^N=1.
 \ee
 To obtain a non-zero answer all $2N$  anticommuting variable must be present  in the integrand,  and in an even permutation of  increasing numerical  order to get a $+1$. Each interchange of adjacent variables gives a factor of $-1$.

 Under linear changes of variables 
 \bea
\psi^\alpha\to \psi'^{\alpha}&=&{A^\alpha}_\beta\psi^\beta\nonumber\\
\bar \psi_\alpha \to \bar\psi'_\alpha&=& \bar \psi_\beta {B^\beta}_\alpha 
\eea
we have 
\bea
d[\psi]\to [d\psi'] &=& {\rm det}[A]^{-1} d[\psi],\nonumber\\
d[\bar \psi]\to d[\bar\psi']&=& d[\bar \psi] \,{\rm det}[B]^{-1}
\eea
in which the jacobian factors are  the  {\it inverse\/} of the commuting variable version.

 If  ${L}$, with entries ${L^\alpha}_\beta$, is an $N$-by-$N$ matrix representing a  linear map $L:V\to V$, we expand the exponential function in the first line below and use the definition to  get  
\bea
Z(L)&=&\int [d\bar \psi d \psi] \exp\{\bar \psi_\alpha {L^\alpha}_\beta \psi^\beta\}, \nonumber\\
&=& \frac 1 {N!} \epsilon_{\alpha_1\ldots \alpha_N}\epsilon^{\beta_1\ldots \beta_N}{L^{\alpha_1}}_{\beta_1}\cdots {L^{\alpha_N}}_{\beta_N}\nonumber\\
&=& \phantom{\int}\!\!\!{\rm det\,}[{L}].
\eea
The integral for the two-variable correlator or propagtor 
\bea
\vev{\bar\psi_\rho \psi^\sigma} &\stackrel{\rm def}{=} & \frac{1}{Z({L})} \int [d\bar \psi]  [d \psi] \bar\psi_\rho \psi^\sigma \exp\{\bar \psi_\alpha {L^\alpha}_\beta \psi^\beta\}, \nonumber\\
&=& {[{L}^{-1}]^\sigma}_\rho
\eea
follows because the explicit $\bar\psi_\rho \psi^\sigma$ factor forces the omission of the term containing
 ${L^\sigma}_\rho$   in the expansion of the exponential, and from the formula for the inverse of a matrix 
\bea
{L}^{-1}= \frac 1{ {\rm det\,}[{ L}]} {\rm Adj}[{L}],
\eea
where ${\rm Adj}[{L}]$ is the adjugate matrix {\it i.e.\/}\  the transposed matrix of the co-factors. We can check the sign and index placement by observing that the claimed expression gives
\be
\vev{\bar\psi_\alpha  {L^\alpha}_\beta \psi^\beta} = {L^\alpha}_\beta {[L^{-1}]^\beta}_\alpha= \tr \{{\mathbb I}_N\}=N,
\ee
which is correct because inserting an  explicit factor of $\bar\psi_\alpha  {L^\alpha}_\beta \psi^\beta$ into the integral  means that  we need to expand the exponential only to  order $N-1$  to get all the $\psi$'s, and hence we get $N!/(N-1)!=N$ times the integral without the explicit  factor. 

Linear maps  are naturally associated with eigenvectors and and eigenvalues. When  $L$ is diagonalizable ---  {\it i.e.\/}\   possesses  sufficient  eigenvectors $u_n$ to form a basis  --- the determinant is the product of the eigenvalues 
\be
{\rm det}[ L]= \prod_{n=1}^{N} \lambda_n.
\ee
 We can extract  formula   from the integral by diagonalizing $L\to A^{-1} L A= {\rm diag}(\lambda_1,\ldots,\lambda_n)$ and then by  setting $B=A^{-1}$ in the change of variables formul\ae\ given above.
 
Again  when $L:V\to V$ is diagonalizable, we can express the inverse ${[L^{-1}]^\alpha}_\beta$ in terms of the left and right eigenvectors  
 \bea
 Lu_n&=&\lambda_n u_n,\nonumber\\
 v_n^T L&=& \lambda_n v_n^T,
 \label{EQ:left-right-eigenvector}
 \eea
 where $u_n\in V$, $v_n^T \in V^*$.  We have not distinguished between the  left  and right eigenvalues because   the two sets of eigenvalues   coincide:  they are determined by the same characteristic equation. 
For a non-hermitian matrix    a  right (left) eigenvector with eigenvalue $\lambda_n$ is  no longer orthogonal to all right  (left) eigenvectors with a different eigenvalue $\lambda_m$, but it remains true that  a {\it left\/} eigenvector with eigenvalue $\lambda_n$ is orthogonal to every  {\it right} eigenvector with a different eigenvalue $\lambda_m$. 
This is because 
\be
\lambda_n {v}_n^T {u}_m= ({ v}_n^T  {L}){ u}_m={v}_n^T ( {L}{ u}_m) = \lambda_m  { v}_n^T { u}_m,
\ee
so
\be
0=(\lambda_n-\lambda_m)  { v}_n^T {u}_m.
\ee
We may   choose the phases and normalization of the two eigenfunction sets so that 
\be
v^T_nu_m \equiv {v}_{n\alpha} {u}^\alpha_m= \delta_{mn},
 \ee
 and, as   the eigenvectors are being assumed complete, we have
 \be
 \sum_n u_n^\alpha v_{n\beta}= \delta^\alpha_\beta.
 \ee 
 The two sets of eigenvectors $u_n$ and $v^T_n$ then compose  mutually dual bases for $V$ and $V^*$ respectively, and 
 \be
 {[L^{-1}]^\alpha}_\beta = \sum_n \frac 1{\lambda_n} u^\alpha_n v_{n\beta}.
 \ee 
 
 We now naturally expand 
 \be
 \psi^\alpha= \sum_n \chi_n u_n^\alpha, \quad\bar\psi_\beta = \sum_n \bar\chi_n v_{n\beta},
  \ee
  where $\chi_n$, $\bar\chi_n$ are Grassmann variables.

\subsection{Continuous fields}
\label{SEC:continuum} 

Now  consider how these finite integrals work in the continuum where we have an infinite set of Grassmann fields $\psi(x)$, one Grassmann variable for each point $x$, and similarly $\bar\psi(x)$.  We will assume that we have chosen our  gamma matrices to be  Hermitian, in which case the $v_{n}$ left eigenvectors of the previous section coincide with the Hermitian  conjugate  $u^\dagger_n$ of the right eigenvectors $u_n$.

 To avoid dealing with continuous spectra, we   will restrict the discussion   a closed (compact without boundary)   $d$-dimensional spin manifold on which the skew-adjoint Dirac operator 
\be
{\fsl D}= \gamma^a D_a= \gamma^a e^\mu_a \left(\partial_\mu + \textstyle{\frac 12} \sigma^{bc}\,\omega_{bc\mu}\right)
\ee   possesses  a  complete orthonormal set of c-number spinor eigenfunctions   $u_n(x)$ labeled by $n\in {\mathbb Z}$, and   with the properties 
\be 
{\fsl D}u_n=i\lambda_n u _n, \quad \int d^dx \sqrt{g}\, u_n^\dagger(x)u_m(x)  =\delta_{mn},\quad \sum_n u_n(x) u^\dagger_n(x') ={\mathbb I}\,\delta_g^d(x-x').
\ee
Here the the $\lambda_n$ are real, $\mathbb I$ is the identity matrix in spinor space,  and the distribution $\delta_g^d(x-x')$ obeys 
\be
\int d^dx \sqrt{g} \,\delta^d(x-y)=1.
\ee

In Euclidean signature  there is no preferred ``$\gamma^0$'' and therefore no inherent need to distinguish between $\psi^\dagger(x)$ and $\bar\psi(x)$, but when we  
 use the eigenmodes to expand out the Grassmann-valued Fermi fields it is convenient to write 
\bea
\psi(x)&=& \sum_n u_n(x) \chi_n,\nonumber\\
\bar\psi(x)&=& \sum_n u^\dagger_n(x) \bar\chi_n.
\eea
As before, the    Grassmann variables $\bar\chi_n$ and $\chi_n$ are independent, and {\it not} related by any notion of complex conjugation, but  when   $\overline {(\ldots)}$ is applied to an expression containing  $\psi(x)$ we understand that  it not only transposes and complex conjugates matrices and  the spinor functions $u_n(x)$ but it also changes any   $\chi_n$'s  into   $\bar \chi_n$'s.

The Euclidean action functional for the Dirac field can therefore be taken as
\be
S[\psi, \bar \psi]= \int d^dx\sqrt{g} \left\{ {\textstyle \frac 12}\left(\bar \psi ({\fsl D}\psi)-\overline{({\fsl D}\psi)}\psi\right)+m \psi^\dagger \psi \right\}.
\ee
Equivalently
\be
S[\psi, \bar \psi]= \int d^dx\sqrt{g} \left\{ {\textstyle \frac 12}\left(\bar \psi {\gamma^a (D_a}\psi)- (D_a \bar \psi)\gamma^a  \psi\right)+m \psi^\dagger \psi \right\},
\ee
where the covariant derivative $D_a$ acting on conjugate spinors $\psi^\dagger$ or $\bar\psi$ is
\be
D_a \bar \psi = e^\mu_a \bar \psi\left(\overleftarrow \partial_\mu -  \textstyle{\frac 12} \sigma^{bc}\,\omega_{bc\mu}\right)
\ee
with $\bar \psi  \overleftarrow \partial_\mu= \partial_\mu \bar \psi$.
The second   form has the advantage of treating $\psi$ and $\bar \psi$ symmetrically.

On inserting the eigenfunction expansions and using the eigenfunction orthonormality to evaluate the space-time integrals, the   Euclidean action functional becomes diagonal
 \bea
 S[\psi,\bar \psi]&=& \int d^dx\sqrt{g} \left\{ {\textstyle \frac 12}\left(\bar \psi ({\fsl D}\psi)-\overline{({\fsl D}\psi)} \psi\right)+m \bar \psi \psi \right\}\nonumber \\
&=& \sum_n (i\lambda_n +m)\bar\chi_n \chi_n.
\eea
The vacuum-amplitude partition  function is   now  formally given by the   Berezin integral  
\bea
Z&=& \int d[\bar\psi]d[\psi] \exp\{S[\psi,\psi^\dagger]\}\nonumber\\
&=&\int \prod_n d[\bar \chi_n] d[\chi_n] \exp\{  \sum_n (i\lambda_n +m)\bar\chi_n \chi_n \}\nonumber\\
&=& \prod_n(i\lambda_n +m)\nonumber\\
&=& {\rm Det}({\fsl D}+m).
\eea
Here ${\rm Det}({\fsl D}+m)$ is the Matthews-Salam functional determinant \cite{matthews-salam}   The infinite product  over the eigenvalues  usually  needs  some form of regularization. 

In the path integral  the   Berezinian  version of the jacobian  determinants involved in the  change of integration measure  from $d [\bar\psi(x)]d[\psi(x)]$ to $ d[\bar \chi_n] d[\chi_n]$ cancel one another just as they do in the finite case. We are, in effect, performing a unitary similarity transformation 
$$
({\fsl D}+m)= U^\dagger {\rm diag}(i\lambda_n+m) U
$$
in which  ${\rm Det}(U)= [{\rm Det} (U^\dagger)]^{-1}$. This formal cancellation is not affected by some of the  $\lambda_n$ being zero. 
\subsection{Majorana fermions: determinants {\it vs\/}.\ Pfaffians}
\label{SEC:pfaffian}

For Majorana fermions we  require an  integral containing a    skew symmetric matrix ${Q}_{ij}$ representing a skew bilinear (symplectic)  form $Q:V\times V\to {\mathbb C}$. As the matrix $Q$ is equipped with  two lower indices,  we no longer need distinguish between  $\psi^\alpha$ and $\bar \psi_\alpha$. For a  $2N$-by-$2N$  matrix we  have    $\psi^\alpha$, $\alpha=1,\dots,2N$, and  the defining   integral becomes  
\be
\int [d\psi] \psi^1\cdots \psi^{2N}=1.
\ee
Again all $\psi^\alpha$ must be present and in numerical order to get $+1$. Thus
\be
\int [d\psi] \psi^{\alpha_1} \cdots \psi^{\alpha_{2N}}= \epsilon^{\alpha_1\ldots \alpha_{2N}}.
\ee

Using this definition we  evaluate
\bea
Z({Q})&=&  \int [d \psi] \exp\left\{ \frac 12 \psi^\alpha Q_{\alpha\beta} \psi^\beta\right\}
 \nonumber\\
&=&\frac 1{2^N N!} \epsilon^{\alpha_1\ldots \alpha_{2N}} Q_{\alpha_1\alpha_2}\cdots 
Q_{\alpha_{2N-1}\alpha_{2N}}\nonumber\\
&\equiv&\phantom{\int}\!\!\!{\rm Pf\,} [{ Q}].
\eea
The last two lines serve to  define the Pfaffian of the skew symmetric matrix ${Q}$.
The two-variable correlator is now 
\bea
\vev{\psi^\rho\psi^\sigma}&=& \frac 1 {Z({Q})}  \int [d \psi]\, \psi^\rho\psi^\sigma \exp\left\{ \frac 12 \psi^\alpha Q_{\alpha\beta} \psi^\beta\right\}\nonumber\\
&=& [Q^{-1}]^{\sigma\rho}.
\eea
Again we check the sign and index placement by computing
\be
\frac 12 \vev{\psi^\rho Q_{\rho\sigma} \psi^\sigma}=\frac 12 Q_{\rho\sigma}[Q^{-1}]^{\sigma\rho} =\frac 12 \tr\{ \mathbb I_{2N}\}=N.
\ee

Regarding ${Q}$ simply as  a numerical matrix  there is a well-known identity 
\be
({\rm Pf\,} {Q})^2= {\rm det\,} [{Q}],
\ee
  which implies that the  Pfaffian of a matrix  is  a square-root of its  determinant. We need to interpret this statement with care.   A linear map  $L$ and  a symplectic   form $Q$  are  rather different mathematical objects. A   linear map $L:V\to V$ possesses  eigenvalues and eigenvectors while  a  bilinear form  $Q:V\times V\to {\mathbb C}$ does not. The  placement of the indices on their  entries indicates that  their matrix representatives   respond differently to  a change of basis in the vector space $V$:
\bea
{L}&\to& {B}^{-1}{L}{B}, \quad \hbox{(Similarity transformation)},\nonumber\\
{Q}&\to& {B}^T {Q}{B}, \quad\,\,\hbox{(Congruence transformation)}.
\eea
From  
 \be
 {\rm det\,} [ {B}^T {Q}{B} ] = {\rm det\,} [{Q}]\,{\rm det\,} [{B}]^2
 \ee  
 we see that a bilinear  form does not possess  a basis-independent determinant.
 We will show later that 
\be
{\rm Pf} [{B}^T{QB}]=  {\rm Pf}[{Q}] \,{\rm det}[{B}],
\ee 
so a skew bilinear  form does not possess a basis-independent Pffafian.  

To convert a   linear map  into a bilinear form, or {\it vice-versa\/},  we need to have some sort of ``metric" to  lower or raise the first index on  ${L^\alpha}_\beta$ or $Q_{\alpha\beta}$.  For relativistic Majorana fermions in spacetime dimensions 2, 3, 4 (mod 8) this role is   played by the antisymmetric  charge-conjugation matrix ${\mathcal C}_{\alpha\beta}$ and its inverse $[C^{-1}]^{\alpha\beta}$. For pseudo-Majorana fermions in spacetime dimensions 0, 1, 2 (mod 8) the role is  taken  by the  symmetric  time-reversal matrix ${\mathcal T}_{\alpha\beta}$ and its inverse $[T^{-1}]^{\alpha\beta}$.  For symplectic Majorana's we need to include an $\epsilon^{ab}$ acting on the labels $a,b=$ 1, 2  distinguishing the  fermion pair.

The  cases are different.  In the finite dimensional version of the  first we  are given a $2N$-by-$2N$ non-degenerate  skew-symmetric matrix ${C}$ with entries $C_{ij}$ together with   a self-adjoint  linear operator represented by a   $2N$-by-$2N$ hermitian matrix ${L}$ with entries ${L^i}_j$,   such that their product $Q_{ij} =C_{ik}{L^k}_j$ is skew symmetric. The simplest case is
\be
\left[\matrix{0&1\cr -1&0}\right]\left[\matrix{\lambda &0\cr 0&\lambda}\right]= \left[\matrix{0 &\lambda \cr -\lambda &0}\right].
\ee
We wish to   evaluate  ${\rm Pf}[{Q}]={\rm Pf}[{CL}]$ in terms of the eigenvalues of $L$. Each eigenvalue occurs twice and,  after some algebra that we will display later, we find that we need only one of the pair in the result
\be
{\rm Pf}[Q]= {\rm Pf}[ C] \prod_{n=1}^N \lambda_n.
\ee

In the second case  $C$ is replaced by a  $2N$-by-$2N$ non-degenerate  {symmetric\/} matrix ${T}$ with entries $T_{ij}$ such that   $Q_{ij} =T_{ik}{L^k}_j$ is  skew symmetric. The simplest example is
\be
\left[\matrix{1&0\cr 0&-1}\right] \left[\matrix{0&\lambda\cr \lambda &0}\right]=\left[\matrix{0&\lambda \cr -\lambda&0}\right].
\ee
We can again  evaluate   ${\rm Pf}[{Q}]={\rm Pf}[{TL}]$ in terms of the eigenvalues of $L$, but the result is more complicated. The eigenvalues occur in $\pm \lambda_n$ pairs, and if we arbitrarily select $\lambda_n$ to the positive eigenvalue, we find  
\be
{\rm Pf}[TL]= \pm  \sqrt{(-1)^N{\rm det}[T]}  \prod_{n=1}^N (-\lambda_n).
\ee
It is not possible to decide what sign to take for the $\pm$ without more information. In the simplest example above, we need  the minus sign if $\lambda$ is positive and the plus sign if $\lambda$ is negative.

The source  of the difference between the $C$ case and the $T$ case  is that after reducing $C$ to a  standard  symplectic form,  the    matrices $B$ in the subsequent normal-form reduction 
\be
Q\to B^TQB= \bigoplus_{n=1}^N \left[\matrix{0&\lambda_n\cr-\lambda_n &0}\right]
\ee
 belong  to ${\rm Sp}(2N)$ and  symplectic matrices are automatically unimodular. After   reducing $T$ to a standard metric  the  matrices $B$ lie in ${\rm SO}(N,N)$ and such orthogonal matrices can have either  $\pm 1$ as their determinant.  The proofs of these Pfaffian formul\ae\  are given below.

\subsection{Pfaffian  to  Determinant}

 We  can almway rewrite the linear operator ``action" $\bar\psi L\psi$ as 
\be
\bar \psi {L} \psi =\frac 12  [\bar\psi,\psi] \left[\matrix{0 & {L}\cr -{ L}^T &0}\right] \left[\matrix{ \bar\psi \cr \psi}\right].
\ee
When  $L$ is $N$-by-$N$ we can now compute  the Pfaffian of the   skew symmetric matrix  and so find that 
\be
{\rm Pf}\left[\matrix{0 & { L}\cr -{L}^T &0}\right]=(-1)^{N(N-1)/2} {\rm det}[ {L}], 
\ee
The sign comes from the need to rearrange the $d\psi$ and $d\bar\psi$  so as  to put all the $d\bar\psi$'s before the $d\psi$'s  instead of in adjacent pairs.  The dependence on $N$  makes this rewriting less useful in infinite dimensions.

\subsection{Proofs of some Pfaffian formul\ae} 
\label{SEC:pf-proof}

$\bm {{\rm Pf}[B^TQB]= {\rm det}[B]{\rm Pf}[Q]}$:  Start from the definition of the Pfaffian 
\be 
{\rm Pf\,} [{ Q}]=\frac 1{2^N N!} \epsilon^{j_1\ldots j_{2N}} Q_{j_1j_2}\cdots 
Q_{j_{2N-1}j_{2N}}
\ee
and recall that 
\be
\epsilon^{j_1,\ldots j_{2N}} {\rm det} [B]= \epsilon^{i_1,\ldots, i_{2N}}B_{j_1i_1}\cdots B_{j_{2N}i_{2N}}.
\ee
Thus 
\bea
{\rm Pf\,} [B^T QB]&=& \frac 1{2^N N!} \epsilon^{i_1\ldots i_{2N}} [B^TQB]_{i_1i_2}\cdots 
B^TQB]_{i_{2N-1}i_{2N}}\nonumber\\
&=&  
\frac 1{2^N N!} \epsilon^{i_1\ldots i_{2N}} B_{j_1i_1}Q_{j_1j_2}B_{j_2 i_2}\cdots 
B_{j_{2N-1}i_{2N-1}} Q_{j_{2N-1} j_{2N}} B_{j_{2N}i_{2N}}\nonumber\\
&=& \frac 1{2^N N!}\epsilon^{j_1\ldots j_{2N}} {\rm det}[B] Q_{j_1j_2}\cdots Q_{j_{2N-1} j_{2N}}\nonumber\\
&=& {\rm det}[B] {\rm Pf\,} [{ Q}].\quad \blacksquare
\eea

\noindent $\bm {({\rm Pf\,}Q)^2 = {\rm det}[Q]}$: 
Given a $2N$-by-$2N$  non-degenerate skew-symmetric matrix $Q$ with entries in a field, we can repeatedly complete squares to find a  linear map  $B$  that reduces $Q$ to the canonical form\cite{stone-goldbart}  
\be
Q=B^TJB, \quad  J= \bigoplus_1^N\left[\matrix{0&1\cr -1 & 0}\right].
\ee
Taking the Pfaffian of this equation  we get 
 \be
 {\rm Pf}[Q]={\rm det}[B]  {\rm Pf}[ J]={\rm det}[B]. 
 \ee 
Taking the determinant  gives
\be
{\rm det}[Q]= {\rm det}[(B^T J B]=  {\rm det}[B]^2 = ({\rm Pf\,}Q)^2. \quad \blacksquare
\ee

\noindent  $\bm {{\rm Pf}[CL]={\rm Pf}[ C] \prod_{n=1}^N \lambda_n}$: Here    $C$  is skew symmetric and $L$ is a    $2N$-by-$2N$ hermitian matrix such that   $Q_{ij} =C_{ik}{L^k}_i$ is skew symmetric, observe that   the hermiticity  of $L$ implies that $L^*=L^T$, and hence 
$$
Lu_n=\lambda u_n\quad \Rightarrow\quad LC^{-1} u_n^* = \lambda  C^{-1} u_n^*.
$$ 
The skew symmetry of $C^{-1}$ guarantees that $u_n$ and $C^{-1}u^*_n$ are mutually orthogonal
$$
u_n^\dagger (C^{-1} u_n^*)= u_n^{*T} C^{-1}u^*_n=0, 
$$ 
so each eigenvalue of $L$ is therefore doubly degenerate and we can assume that $u_n$ and $C^{-1} u^*_n$, $n=1,\ldots, N$,  together constitute a complete orthonormal set. Let us introduce vectors $\tilde X=(\tilde x_1,\tilde y_1,\cdots \tilde x_n,\tilde y_n)$ and $X=(x_1,y_1,\cdots x_n,y_n)$
 Using the orthonormality we have 
 \bea
\tilde X^TB^T (CL) BX&=& \sum_n (\tilde x_n u_n+\tilde y_n C^{-1} u_n^*)^T CD(x_n u_n+y_nC^{-1} u_n^*)
\nonumber\\
 &=& 
 \sum_n(\tilde x_n u^T_n-\tilde y_n  u_n^\dagger C^{-1} ) CD(x_n u_n+y_nC^{-1} u_n^*)\nonumber\\
 &=&\sum_n\lambda_n (\tilde x_n y_n- \tilde y_nx_n) \nonumber\\
 &=& \tilde X^T \Lambda  X.\nonumber
\eea 
Here 
$$
\Lambda= \bigoplus_{n=1}^N  \left[\matrix{0, &\lambda_n \cr -\lambda_n &0}\right] ,
$$
and ${B}$ is the  $2N$-by-$2N$  matrix 
$$
{B}=[u_1, C^{-1}u_1^*, \cdots , u_n ,C^{-1}u_n^*].
$$ 
We have reduced $CL$ to a canonical form, and taking the Pfaffian we have 
$$
{\rm Pf}[B^T (CL) B] = {\rm Pf}[CL] {\rm det}[B] =\prod_n \lambda_n. 
$$
We need to find an expression for ${\rm det}[B] $.  To do this replace  $L$ by ${\mathbb I}_{2N} $ while keeping the $u_n$ unchanged.  This results in    
$$
B^T C B = J=  \bigoplus_{n=1}^N  \left[\matrix{0 &1 \cr -1 &0}\right].
$$ 
But  ${\rm Pf}[J]=1$ so   ${\rm det}[B]= {\rm Pf}[C]^{-1}$. 
The end result is that 
$$
 {\rm Pf}[CL] = {\rm Pf}[ C] \prod_n \lambda_n. \quad \blacksquare
 $$

\noindent $\bm {{\rm Pf}[{TL}]=\pm  \sqrt{(-1)^N{\rm det}[T]}  \prod_{n=1}^N (-\lambda_n)}$: We are given   $2N$-by-$2N$ non-degenerate  {symmetric\/} matrix ${T}$ with entries $T_{ij}$ such that   $Q_{ij} =T_{ik}{L^k}_i$ is  skew symmetric. An  example to bear in mind is
$$
\left[\matrix{1&0\cr 0&-1}\right] \left[\matrix{0&\lambda\cr \lambda &0}\right]=\left[\matrix{0&\lambda \cr -\lambda&0}\right].
$$
To  evaluate  ${\rm Pf}[{Q}]={\rm Pf}[{TL}]$ in terms of the eigenvalues of $L$ we use a similar strategy as before. In this case the hermiticity  of $L$ gives us 
$$
Lu_n=\lambda_n u_n\quad \Rightarrow\quad LT^{-1} u_n^* =- \lambda  T^{-1} u_n^*,
$$ 
and if $\lambda_n$ is non zero $T^{-1} u_n^*$  is orthogonal to $u_n$ because they have different eigenvalues.
 The  non-zero-mode eigenvectors of  $L$   therefore come in opposite eigenvalue pairs. If there are no zero modes  the   $u_n$ and $T^{-1} u^*_n$, $n=1,\ldots, N$,  together constitute a complete orthonormal set. We will  take $\lambda_n$ to be the positive eigenvalue.
 Again set  $\tilde X=(\tilde x_1,\tilde y_1,\cdots \tilde x_n,\tilde y_n)$ and $X=(x_1,y_1,\cdots x_n,y_n)$
 and use the orthonormality and symmetry of $T^{-1}$  to conclude that  
 \bea
\tilde X^TB^T (TL) BX&=& \sum_n (\tilde x_n u_n+\tilde y_n T^{-1} u_n^*)^T TL(x_n u_n+y_nT^{-1} u_n^*)
\nonumber\\
 &=& 
 \sum_n(\tilde x_n u^T_n+\tilde y_n  u_n^\dagger T^{-1} ) TL(x_n u_n+y_nT^{-1} u_n^*)\nonumber\\
 &=&\sum_n(-\lambda_n) (\tilde x_n y_n- \tilde y_nx_n) \nonumber\\
 &=& \tilde X^T(- \Lambda)  X\nonumber
\eea 
where 
$$
\Lambda= \bigoplus_{n=1}^N  \left[\matrix{0, &\lambda_n \cr -\lambda_n &0}\right] ,
$$
and ${B}$ is the  $2N$-by-$2N$  matrix 
$$
{B}=[u_1, T^{-1}u_1^*, \cdots , u_n ,T^{-1}u_n^*]
$$ 
Thus 
$$
{\rm Pf}[B^T (TL) B] = {\rm Pf}[TL] {\rm det}[B] =\prod_n (-\lambda_n). 
$$
In this case, however we cannot take the Pfaffian after replacing $L$ by ${\mathbb I}$ because $T$ is not skew symmetric. We can still replace $L\to {\mathbb I}$ and find that
$$ 
{\tilde X} B^T TB  X= \sum_n (\tilde x_n y_n+ \tilde y_n x_n)= \tilde XG X
$$
where 
$$
G= \bigoplus_{n=1}^N  \left[\matrix{0, &1\cr 1&0}\right]. 
$$
We can now take the  determinant  to conclude that
$$
{\rm det}[B^T T B] = {\rm det}[G]= (-1)^N
$$
 and so ${\rm det}[B]^2{\rm det T} =(-1)^N$. Hence
 $$
{\rm Pf}[TL]= \pm \sqrt{(-1^N){\rm det}[T]}  \prod_n(- \lambda_n),
$$  
where the $\pm$ sign comes from the need to take $\sqrt{{\rm det}[B]^2}$.
The uncertainty as to which root to take is inevitable. We arbitrarily assigned $u_n$ to the positive eigenvalue  rather than to the negative. If we make the opposite  choice for some  eigenvalue pair, the product formula must change sign whilst  ${\rm Pf}[Q]$ itself is indifferent  to our choice. $\blacksquare$


\section{Canonical forms for complex matrices}
\label{SEC:canonical}

 Here are some lesser known, but useful reductions of matrices with complex entries:  

If $M$ is an $n$-by-$n$  complex symmetric matrix, then there exists
\cite{autonne,takagi} a unitary matrix $\Omega$  such that 
\be
\Omega^T M \Omega= {\rm diag}(m_1,\ldots, m_n) 
\ee
where the numbers $m_i$ are real and non-negative. This result is useful for diagonalizing symmetric ${\mathcal C}$ and ${\mathcal T}$ matrices and also for Majorana-mass matrices.

\noindent{\bf Proof\/}: the matrix $N = M^\dagger M$  is Hermitian and non-negative, so there is a unitary matrix $V$ such that $V^\dagger N V$  is diagonal with non-negative real entries. Thus $C = V^TM V$ is complex symmetric with $C^\dagger C\equiv V^\dagger N V$   real. Writing $C = X + iY$ with $X$ and $Y$ real symmetric matrices, we have $C^\dagger C = X^2 + Y^2 + i[X,Y ]$. As this expression is real, the commutator must vanish. Because   $X$ and $Y$ commute, there is a real orthogonal matrix $W$ such that both $WXW^T$ and $WYW^T$ are simultaneously diagonal. Set $U = WV^T$ then $U$ is unitary and  the matrix $UMU^T$ is complex diagonal. By post-multiplying $U$ by another diagonal unitary matrix, the diagonal entries can be made to be real and non-negative. Since their squares are the eigenvalues of $M^\dagger M$, they coincide with the singular values of $M$. $\blacksquare$

If $A$ is a complex skew-symmetric matrix, one can use the same strategy to show there exists a unitary matrix $\Omega$ such that\cite{hua} 
$$
\Omega^T A\Omega = \bigoplus_i  \left[\matrix{ 0&\lambda_i \cr -\lambda_i &0}\right]\oplus  {\rm diag}(0,\ldots,0), 
$$
and the $\lambda_i $ are the positive square-roots of the   eigenvalues $\lambda_i^2$  of $A^\dagger A$. We can use this to show that the Pfaffian of a skew matrix $Q$ is the product of the square roots of the eigenvalues of $Q^\dagger Q$, but only {\it up to a phase} equal to  ${\rm det}[\Omega]$ .

\section{The ${\mathsf C}$ and ${\mathsf T}$ operations in Condensed Matter Systems}
\label{SEC:condensed}

The eightfold Bott periodicity we have uncovered in the Dirac equation manifests itself in the various discrete symmetries of non-relativistic systems \cite{kitaev}.  For a  
 review  see  \cite{ryu-review}.  This setting is useful for explaining why charge conjugations is a \underline{unitary linear} map on the many-particle Hilbert space despite involving  a complex conjugation operation.

Condensed matter physics is usually formulated in Hamiltonian language.
and we  will restrict ourselves to non-interacting Hamiltonians built from  a set of fermion annihilation and creation operators $\Psi_\alpha$ and $\Psi^\dagger_\alpha$ that obey
 \be
 \{\Psi_\alpha,\Psi_\beta\}=0,\quad  \{\Psi_\alpha,\Psi^\dagger_\beta\}= \delta_{\alpha\beta}.
 \ee
 We will need to distinguish between a vacuum state $\ket{\rm empty}$ such that $\Psi_\alpha \ket{\rm empty}=0$ for all $\alpha$, and a ground state $\ket{\rm gnd}$ in which all negative energy states are occupied.

\subsection{${\mathsf C}$ or particle-hole symmetry} Suppose that we have a non-interacting many-fermion  Hamiltonian  
\be
\hat H=\Psi^\dagger_\alpha H_{\alpha\beta} \Psi_\beta
\ee
where the $N$-by-$N$ one-particle Hamiltonian matrix $H_{\alpha\beta}$ is traceless and obeys
\be
C H^* C^{-1}= -H
\ee
 for some unitary matrix $C$. The Bogoliubov-de-Gennes Hamiltonian  for superconducting  systems has this property. Now   
\be
Hu_n=\lambda_n u_n\quad\Rightarrow \quad HCu_n^* = -\lambda_n Cu^*_n,
\ee
so, when $\lambda$ is non zero, the single-particle eigenfunctions come in opposite-eigenvalue  pairs.     In the absence of zero energy states the ground state $\ket{\rm gnd}$ has all negative-energy states occupied and   is non-degenerate.

We define the action of a  {\it unitary\/} particle-hole operator ${\mathfrak C}$ on the many-body Fock space by ${\mathfrak C}\ket{\rm empty}=\ket{\rm empty}$ and 
\be
{\mathfrak  C}\Psi_\beta {\mathfrak  C}^{-1}= \Psi_\alpha^\dagger C_{\alpha\beta}, \quad {\mathfrak C}\Psi^\dagger_\beta {\mathfrak  C}^{-1}= C^\dagger_{\beta\alpha}\Psi_\alpha.
\ee

When  
 ${\mathfrak C}$ acts on the Hamiltonian  we have 
\bea 
{\mathfrak C}\hat H {\mathfrak  C}^{-1}&=&
{\mathfrak  C}\Psi^\dagger_\alpha H_{\alpha\beta} \Psi_\beta {\mathfrak  C}^{-1}\nonumber\\
&=&{\mathfrak C}\Psi^\dagger_\alpha {\mathfrak  C}^{-1}{\mathfrak  C}H_{\alpha\beta}{\mathfrak  C}^{-1}{\mathfrak  C}  \Psi_\beta {\mathfrak  C}^{-1}\nonumber\\
&=& {\mathfrak  C}\Psi^\dagger_\alpha {\mathfrak  C}^{-1}H_{\alpha\beta}{\mathfrak  C} \Psi_\beta {\mathsf C}^{-1}\nonumber\\
&=& C^\dagger_{\alpha\rho}\Psi_{\rho}  H_{\alpha\beta} \Psi^\dagger_\sigma C_{\sigma\beta}\nonumber\\
&=&- \Psi^\dagger_\sigma C_{\sigma\beta} H_{\alpha\beta}C^\dagger_{\alpha\rho}\Psi_{\rho}\nonumber\\
&=&- \Psi^\dagger_\sigma C_{\sigma\beta} H^T_{\beta \alpha}C^\dagger_{\alpha\rho}\Psi_{\rho}\nonumber\\
&=&- \Psi^\dagger_\sigma C_{\sigma\beta} H^*_{\beta \alpha}C^\dagger_{\alpha\rho}\Psi_{\rho}\nonumber\\
&=&+ \Psi^\dagger_\sigma H_{\sigma\rho} \Psi_{\rho}.\nonumber\\
&=& \hat H.
\eea
Thus  the one-particle transformation on $H$ leaves the many-particle Hamiltonian invariant. 

We  used $C^{*\dagger} = C^T$ and the tracelessness (line 5 $\to$ 6) and hermiticity of $H$ in the above manipulations. 
More importantly, and   despite the appearance of the complex conjugation  ``$*$" in   $H^* = -C^{-1}HC$, the many-body operator  ${\mathfrak  C}$ must act  on the Fock space {\it linearly\/}:
\be
{\mathfrak  C}(\lambda |\psi_1\rangle+\mu |\psi_2\rangle)= \lambda {\mathfrak  C}|\psi_1\rangle+\mu {\mathfrak  C}|\psi_2\rangle.
\ee
The linearity is  required  in the step 
\be
{\mathfrak C}H_{\alpha\beta}{\mathfrak  C}^{-1}= H_{\alpha\beta}.
\ee
 
 If we write 
\be
\Psi_\alpha = \sum_n u_{n\alpha} \hat a_n
\ee
with $n>0$ corresponding to positive energy and $n<0$ to negative, then in the absence of zero energy states the  ground state $\ket{\rm gnd}$ is specified up to phase by  
\be
\hat a_n\ket{\rm gnd}= \hat a_{-n}^\dagger \ket{\rm gnd}=0, \quad n>0.
\ee
Let $C^T=\lambda C$, 
 $\lambda=\pm 1$.  From this we deduce that ${\mathfrak  C} a_n{\mathfrak C}^{-1} = \lambda a^\dagger_{-n}$ and hence that     ${\mathfrak  C}\ket{\rm gnd}=\ket{\rm gnd}$. 
 We also have that 
  \be
  {\mathfrak  C}(\Psi^\dagger_\beta \Psi_\beta -N/2){\mathfrak  C}^{-1}= \Psi_\alpha \Psi^\dagger_\alpha -N/2 = - (\Psi^\dagger_\alpha \Psi_\alpha -N/2),
  \ee
  so the sign of the normal-ordered charge operator 
  \be
  \hat Q= \frac 12 (\Psi^\dagger_\beta \Psi_\beta -\Psi_\beta\Psi^\dagger_\beta)=\Psi^\dagger_\beta \Psi_\beta-N/2
  \ee
   is reversed.

\subsection{${\mathsf T}$ and time-reversal}

 Again consider a  non-interacting many-fermion  Hamiltonian  
\be
\hat H=\Psi^\dagger_\alpha H_{\alpha\beta} \Psi_\beta,
\ee
but now assume that the one-particle Hamiltonian matrix $H_{\alpha\beta}$  obeys 
\be
T H^* T^{-1}= +H
\ee
for some unitary matrix $T$.  This condition tells us that  if  $u_n(x,t)$ obeys
\be
\left(i\frac{\partial}{\partial t}-H\right) u_n(x,t)=0
\ee
then $Tu_n^*(x,-t)$ obeys the same equation:
\be
\left(i\frac{\partial}{\partial t}-H\right) Tu^*_n(x,-t)=0.
\ee

We define the action of an  {\it anti-unitary\/} time reversal operator ${\mathfrak T}$ on the many-body Fock space by
\be
{\mathfrak T}\Psi_\beta {\mathfrak T}^{-1}= T^\dagger _{\beta \alpha} \Psi_\alpha , \quad {\mathfrak T}\Psi^\dagger_\beta {\mathfrak T}^{-1}= \Psi^\dagger_\alpha T_{\alpha\beta}.
\ee
Then ${\mathfrak T}\Psi^\dagger _\beta\Psi_\beta {\mathfrak T}^{-1}= \Psi^\dagger _\beta\Psi_\beta$, so the charge is unchanged, and 
\bea 
{\mathfrak T}\hat H {\mathfrak  T}^{-1}&=&
{\mathfrak T}\Psi^\dagger_\alpha H_{\alpha\beta} \Psi_\beta {\mathfrak T}^{-1}\nonumber\\
&=&{\mathfrak T}\Psi^\dagger_\alpha {\mathfrak T}^{-1}{\mathfrak T}H_{\alpha\beta}{\mathfrak T}^{-1}{\mathfrak T}  \Psi_\beta {\mathfrak T}^{-1}\nonumber\\
&=& {\mathfrak T}\Psi^\dagger_\alpha {\mathfrak T}^{-1}H^*_{\alpha\beta}{\mathfrak  T} \Psi_\beta {\mathfrak T}^{-1}\nonumber\\
&=& \Psi_{\rho} T_{\rho\alpha} H_{\alpha\beta}^* T^\dagger_{\beta\sigma} \Psi_\sigma \nonumber\\
&=& \Psi^\dagger_\rho H_{\rho\sigma} \Psi_{\sigma}.\nonumber\\
&=& \hat H.
\eea
Again the  transformation of  $H$ leaves  the many-particle Hamiltonian invariant.
We required   ${\mathfrak  T}$ to be   a {\it anti-linear\/} map
because we need  
\be
{\mathfrak T}H_{\alpha\beta}{\mathfrak T}^{-1}= H^*_{\alpha\beta}.
\ee

\end{document}